\documentclass[
preprintnumbers, 
showpacs, 
amsmath, 
showkeys, 
amssymb, 
aps, 
tightenlines,
superscriptaddress, 
twocolumn,
longbibliography, 
letterpaper,
nofootinbib,
]{revtex4-2}
\usepackage{lipsum, babel}
\usepackage{color}
\usepackage{hyperref}
\usepackage{bm}
\usepackage{dcolumn} % Align table columns on decimal point
\usepackage{booktabs} % Professional-quality tables
\usepackage{amsfonts}
\usepackage{mathrsfs}
\usepackage{graphicx}
\usepackage{amsmath}
\usepackage{float}
\usepackage{braket}
\usepackage{natbib}
\usepackage{orcidlink}

\hypersetup{
	colorlinks=true,
	linkcolor=red,
	citecolor=blue,
}

\begin{document}
	
	\newcommand{\sgn}{\operatorname{sgn}}
	\newcommand{\hhat}[1]{\hat {\hat{#1}}}
	\newcommand{\pslash}[1]{#1\llap{\sl/}}
	\newcommand{\kslash}[1]{\rlap{\sl/}#1}
	\newcommand{\lab}[1]{}
	% to create labels.
	%\newcommand{\iref}[2]{\footnote{\hyperlink{lb:#1}{\textit{$^\spadesuit$#2}}}}
	%\newcommand{\iref}[2]{}
	% reference to other part in this notes.
	\newcommand{\sto}[1]{\begin{center} \textit{#1} \end{center}}
	\newcommand{\rf}[1]{{\color{blue}[\textit{#1}]}}
	% Reference
	\newcommand{\eml}[1]{#1}
	% Emphasis
	\newcommand{\el}[1]{\label{#1}}
	% Equation labeling
	\newcommand{\er}[1]{Eq.\eqref{#1}}
	% Equation Reference
	\newcommand{\df}[1]{\textbf{#1}}
	% Temporarily replace \textbf
	\newcommand{\mdf}[1]{\pmb{#1}}
	% Use for vectors etc.
	\newcommand{\ft}[1]{\footnote{#1}}
	% footnote.
	\newcommand{\n}[1]{$#1$}
	% Use for numbers etc.
	% \newcommand{\cjktext}[1]{\begin{CJK}{GB}{gbsn} #1 \end{CJK}} 
	% Language support
	\newcommand{\fals}[1]{$^\times$ #1}
	% wrong statement
	\newcommand{\new}{{\color{red}$^{NEW}$ }}
	% update
	% \newcommand{\ci}[1]{\cite{#1}}
	\newcommand{\ci}[1]{}
	\newcommand{\de}[1]{{\color{green}\underline{#1}}}
	\newcommand{\ke}{\rangle}
	\newcommand{\br}{\langle}
	\newcommand{\lb}{\left(}
	\newcommand{\rb}{\right)}
	\newcommand{\lbk}{\left[}
	\newcommand{\rbk}{\right]}
	\newcommand{\blb}{\Big(}
	\newcommand{\brb}{\Big)}
	\newcommand{\nn}{\nonumber \\}
	\newcommand{\p}{\partial}
	\newcommand{\pd}[1]{\frac {\partial} {\partial #1}}
	\newcommand{\cd}{\nabla}
	\newcommand{\cc}{$>$}
	% ##### ###### ###### ######
	\newcommand{\bqa}{\begin{eqnarray}}
		\newcommand{\eqa}{\end{eqnarray}}
	\newcommand{\bqe}{\begin{equation}}
		\newcommand{\eqe}{\end{equation}}
	\newcommand{\bay}[1]{\left(\begin{array}{#1}}
		\newcommand{\eay}{\end{array}\right)}
	\newcommand{\eg}{\textit{e.g.} }
	\newcommand{\ie}{\textit{i.e.}, }
	\newcommand{\iv}[1]{{#1}^{-1}}
	\newcommand{\st}[1]{|#1\ke}
	\newcommand{\at}[1]{{\Big|}_{#1}}
	\newcommand{\zt}[1]{\texttt{#1}}
	\newcommand{\non}{\nonumber}
	\newcommand{\m}{\mu}
	% ##### ###### ###### ######
	% Greek Letters
	\def\xa{{m}}
	\def\xA{{m}}
	\def\xb{{\beta}}
	\def\xB{{\Beta}}
	\def\xd{{\delta}}
	\def\xD{{\Delta}}
	\def\xe{{\epsilon}}
	\def\xE{{\Epsilon}}
	\def\xve{{\varepsilon}}
	\def\xg{{\gamma}}
	\def\xG{{\Gamma}}
	\def\xk{{\kappa}}
	\def\xK{{\Kappa}}
	\def\xl{{\lambda}}
	\def\xL{{\Lambda}}
	\def\xo{{\omega}}
	\def\xO{{\Omega}}
	\def\xvp{{\varphi}}
	\def\xs{{\sigma}}
	\def\xS{{\Sigma}}
	\def\xt{{\theta}}
	\def\xvt{{\vartheta}}
	\def\xT{{\Theta}}
	% ##### ###### ###### ######
	\def \Tr {{\rm Tr}}
	\def\CA{{\cal A}}
	\def\CC{{\cal C}}
	\def\CD{{\cal D}}
	\def\CE{{\cal E}}
	\def\CF{{\cal F}}
	\def\CH{{\cal H}}
	\def\CJ{{\cal J}}
	\def\CK{{\cal K}}
	\def\CL{{\cal L}}
	\def\CM{{\cal M}}
	\def\CN{{\cal N}}
	\def\CO{{\cal O}}
	\def\CP{{\cal P}}
	\def\CQ{{\cal Q}}
	\def\CR{{\cal R}}
	\def\CS{{\cal S}}
	\def\CT{{\cal T}}
	\def\CV{{\cal V}}
	\def\CW{{\cal W}}
	\def\CY{{\cal Y}}
	\def\BC{\mathbb{C}}
	\def\BR{\mathbb{R}}
	\def\BZ{\mathbb{Z}}
	\def\sA{\mathscr{A}}
	\def\sB{\mathscr{B}}
	\def\sF{\mathscr{F}}
	\def\sG{\mathscr{G}}
	\def\sH{\mathscr{H}}
	\def\sJ{\mathscr{J}}
	\def\sL{\mathscr{L}}
	\def\sM{\mathscr{M}}
	\def\sN{\mathscr{N}}
	\def\sO{\mathscr{O}}
	\def\sP{\mathscr{P}}
	\def\sR{\mathscr{R}}
	\def\sQ{\mathscr{Q}}
	\def\sS{\mathscr{S}}
	\def\sX{\mathscr{X}}
	
	% 你的 ORCID 链接

	% 使用命令插入 ORCID 链接

	\author{Wei Kou\orcidlink{0000-0002-4152-2150}}
\email{kouwei@impcas.ac.cn}
\affiliation{Institute of Modern Physics, Chinese Academy of Sciences, Lanzhou 730000, Gansu Province, China}
\affiliation{Southern Center for Nuclear Science Theory (SCNT), Institute of Modern Physics, Chinese Academy of Sciences, Huizhou 516000, Guangdong Province, China}
\affiliation{School of Nuclear Science and Technology, University of Chinese Academy of Sciences, Beijing 100049, China}
\affiliation{State Key Laboratory of Heavy Ion Science and Technology, Institute of Modern Physics, Chinese Academy of Sciences, Lanzhou 730000, Gansu Province, China}

\author{Xurong Chen}
\email{xchen@impcas.ac.cn}
\affiliation{Institute of Modern Physics, Chinese Academy of Sciences, Lanzhou 730000, Gansu Province, China}
\affiliation{Southern Center for Nuclear Science Theory (SCNT), Institute of Modern Physics, Chinese Academy of Sciences, Huizhou 516000, Guangdong Province, China}
\affiliation{School of Nuclear Science and Technology, University of Chinese Academy of Sciences, Beijing 100049, China}
\affiliation{State Key Laboratory of Heavy Ion Science and Technology, Institute of Modern Physics, Chinese Academy of Sciences, Lanzhou 730000, Gansu Province, China}

	\title{Probing Saturon-like Limits in QCD Systems}
	%\date{\today}

	\begin{abstract}

High-occupancy QCD matter enters a saturated regime when its entropy or occupancy approaches the unitarity bound $\sim 1/\alpha$, the “saturon” criterion. We test this criterion for protons and nuclei at small $x$ using analytic and numerical solutions of the BK equation. From these solutions we construct the gluon occupancy $N_g(x)$ and a thermodynamic entropy $S(x)$ via an Unruh-like temperature $T = Q_s/(2\pi)$ and an emergent gluon mass $M_g \sim Q_s$. For protons, both $N_g$ and $S$ rise toward small $x$ yet stay below $1/\alpha_s$ in our baseline setup. For nuclei, by contrast, the nuclear entropy $S_A$ attains the $1/\alpha_s$ benchmark in a small-$x$ window where the proton does not. This singles out nuclei as the natural environment to search for saturon-like behavior and motivates precision small-$x$ measurements and high-occupancy $pA$ and $AA$ collisions.

	\end{abstract}

	%\pacs{04.60.Bc, 04.62.+v, 04.70.Dy}
	
	\maketitle
	
	\section{Introduction}
	\label{sec:introduction}
	In high-energy quantum field theory, an emerging concept has stimulated renewed consideration of saturation, entropy, and semiclassical behavior: the saturon. First introduced by Dvali et al. \cite{Dvali:2019jjw,Dvali:2019ulr,Dvali:2020wqi,Dvali:2021jto}, a saturon refers to a highly occupied composite field configuration that attains maximal entropy under the constraint of unitarity. The thermodynamic properties of saturons, such as entropy-area analogies and thermal-like behavior, evoke parallels with black holes, although they arise within non-gravitational and renormalizable theories. When the occupation number n and the coupling constant $\alpha$ satisfy $n\alpha \sim \mathcal{O}(1)$, the system exhibits semiclassical behavior, and the entropy scales as $S \sim 1/\alpha$, which in four dimensions resembles the Bekenstein-Hawking entropy relation $S \sim R^2$ \cite{Bekenstein:1973ur,Hawking:1975vcx}.
	
	Moreover, saturons are not merely theoretical constructs: because their entropy reaches its maximal value, they can be produced in a thermal bath through quantum transitions without exponential suppression. This contrasts with conventional topological solutions or black hole formation mechanisms, which are typically subject to exponential suppression by a Boltzmann factor or dynamical constraints. Such entropy-driven enhancement makes saturons compelling candidates for dark matter and suggests that they may also yield observable signatures in high-energy collisions \cite{Dvali:2023xfz}. 
	
	The potential connection between the infrared behavior of the strong interaction and gravity has long been a subject of intense discussion. One of the most prominent manifestations of this idea is the application of the Anti-de Sitter/Conformal Field Theory (AdS/CFT) correspondence in strongly interacting theories \cite{tHooft:1973alw,Polyakov:1981rd,Maldacena:1997re,Aharony:1999ti}. The rapid development of this framework has led to new interpretations in black hole thermodynamics, quantum field theory, and hadronic physics, particularly by relating certain black hole properties to entanglement and other features emerging in strong interaction studies \cite{Dvali:2021ooc,Kou:2022dkw} (See others works related entropy \cite{Kutak:2011rb,Kutak:2023cwg,Caputa:2024xkp,Chachamis:2023omp,Hentschinski:2024gaa,Hatta:2024lbw,Dumitru:2023qee,Ramos:2022gia,Moriggi:2024tbr,Ramos:2020kaj,Peschanski:2012cw,Armesto:2019mna,Neill:2018uqw,Kovner:2018rbf,Chachamis:2023omp,Liu:2022bru,Liu:2022hto,Liu:2022qqf,Liu:2023eve,Stoffers:2012mn,Asadi:2023bat}.). As the fundamental theory of strong interactions, QCD provides a basis for exploring hadron structure and nuclear matter properties. Its infrared dynamics can also be investigated through the mapping of four-dimensional Minkowski field theories onto the boundary of a five-dimensional AdS spacetime.
    Distinct from the holographic duality, another profound connection between gauge theories and gravity is found in the Bern-Carrasco-Johansson (BCJ) ``double copy" construction \cite{Bern:2008qj}. While AdS/CFT relates a gauge theory to a gravity theory in a higher-dimensional space, the double copy framework establishes a direct relationship between scattering amplitudes in gauge theories and gravity within the same spacetime dimension, loosely summarized as ``gravity = gauge $\times$ gauge".

    Regardless of the specific framework, black holes remain a focal point for understanding high-density gravitational states. Intriguingly, it has been proposed that the physics of black holes---specifically their saturation of entropy bounds---may share universal features with other high-occupancy systems. Motivated by such universality classes, recent works \cite{Dvali:2021ooc} have suggested a duality between the highly occupied graviton states in black holes and the saturated gluon states in the Color Glass Condensate (CGC) effective theory \cite{Iancu:2002xk,Gelis:2010nm}.
    % Its infrared dynamics can also be investigated through the mapping of four-dimensional Minkowski field theories onto the boundary of a five-dimensional AdS spacetime. A notable recent example of a concrete connection between gravity and QCD is the so-called BCJ ``double copy" relation \cite{Bern:2008qj}, where gravitational perturbative amplitudes are constructed from QCD amplitudes supplemented by additional kinematic factors. As a distinctive subject of gravitational studies, black holes exhibit certain properties that may find correspondence in investigations of the strong interaction. As discussed in Ref. \cite{Dvali:2021ooc}, it is possible to relate the behavior of systems near saturation to both highly occupied graviton states characteristic of black holes and the framework of the Color Glass Condensate (CGC) effective theory \cite{Iancu:2002xk,Gelis:2010nm}.
	
	The central focus of this work is to explore saturation phenomena from the perspective of proton structure. It is well established that in high-energy hadrons, such as protons, the number of gluons increases rapidly as the Bjorken-$x$--the longitudinal momentum fraction carried by gluons--becomes very small. However, when the gluon density reaches a sufficiently high level, nonlinear interactions among gluons, such as gluon-gluon recombination, become significant and inhibit the unbounded growth of gluon density. This phenomenon is referred to as gluon saturation. In other words, the gluon population cannot grow indefinitely without violating unitarity. A central question addressed in this work is whether a system constrained by unitarity, such as the saturated gluon system in a proton, can be connected to the concept of a saturon. This concept, introduced by Dvali et al., refers to highly occupied field configurations that reach maximal entropy while respecting unitarity.
    In Sec. \ref{sec:model}, we offer a concise review of QCD’s nonlinear small-$x$ evolution equations and the phenomenon of gluon saturation within the proton. To assess the saturon status of the proton, we adopt the thermodynamic entropy definition for gluonic systems as originally proposed by Kutak in Ref. \cite{Kutak:2011rb}. Our work focuses on applying this entropy measure—along with the gluon occupation number derived from the Balitsky-Kovchegov (BK) equation—to the specific case of proton and nuclear targets.In Sec. \ref{sec:diss}, we present our computed observables and examine the relationship between high-gluon-occupancy protons and saturon behavior. Finally, we provide a summary and outlook. This study introduces a novel perspective on saturation in the proton system, suggesting that, near the saturation threshold, universal behavior can emerge independently of detailed microdynamics or initial conditions. Such a proton state represents a saturated condition comprised of highly occupied soft quanta, sufficient to form a cross section constrained by unitarity \cite{Dvali:2020wqi,Dvali:2021jto,Dvali:2021ooc}.
	
	\section{Formalism}
	\label{sec:model}
	
	\subsection{Gluon saturation and small-$x$ evolution}
	A prominent feature of QCD in the high-energy regime is the emergence of high-density partons distributions, a consequence dictated by first principles. Due to the presence of the three-gluon interaction term, radiated gluons can further split into additional gluon fields. The evolution of parton distributions with respect to the photon virtuality $Q^2$ and the Bjorken scaling variable $x$ is governed by evolution equations, primarily the Dokshitzer-Gribov-Lipatov-Altarelli-Parisi (DGLAP) equation \cite{Dokshitzer:1977sg,Gribov:1972ri,Gribov:1972rt,Altarelli:1977zs} and the Balitsky-Fadin-Kuraev-Lipatov (BFKL) equation \cite{Balitsky:1978ic,Kuraev:1977fs}. The DGLAP equation accurately describes the QCD dynamics of the proton at moderately large $x$ and intermediate to high $Q^2\gg \Lambda_\mathrm{QCD}^2$, while the BFKL equation complements this by capturing the evolution behavior in the small-$x$ regime, where parton densities become significantly enhanced.
	
\begin{figure*}[htpb]
		\begin{center}
			\includegraphics[width=0.17\textwidth]{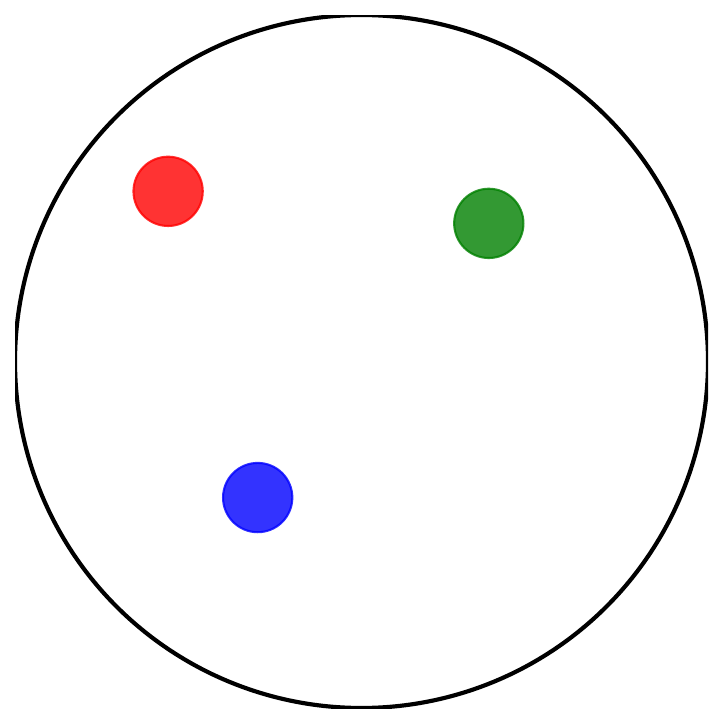}\quad\quad
			\includegraphics[width=0.17\textwidth]{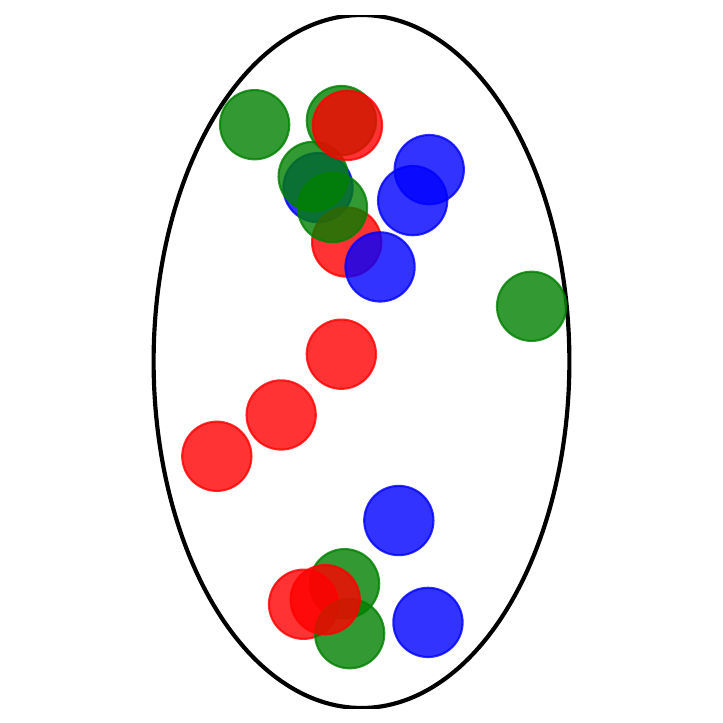}
			\includegraphics[width=0.17\textwidth]{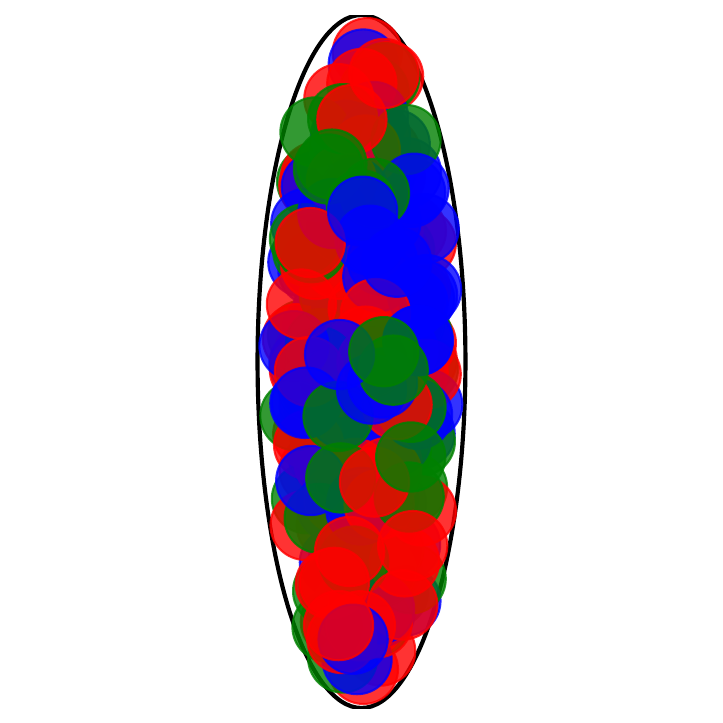}
		\end{center}
		\caption{The gluon density inside the proton increases from low to high energies (from left to right), eventually reaching a state of gluon saturation as a result of overlap and gluon recombination.}
		\label{fig:saturation}
	\end{figure*}

	The unbounded growth of high-energy gluons violates unitarity. In practice, as the number of gluons increasingly populates the phase space of the proton, spatial overlap among them becomes inevitable. When the overlapping gluons come sufficiently close, gluon recombination occurs, leading to saturation (See FIG. \ref{fig:saturation}). This nonlinear effect is not accounted for within the framework of the BFKL equation. In fact, the earliest nonlinear corrections to the DGLAP equation were introduced in Refs. \cite{Gribov:1983ivg,Mueller:1985wy} and are known as the GLR-MQ equation. Subsequently, the Jalilian-Marian-Iancu-McLerran-Weigert-Leonidov-Kovner (JIMWLK) equation \cite{Jalilian-Marian:1997jhx,Jalilian-Marian:1997ubg,Iancu:2000hn,Iancu:2001ad,Weigert:2000gi} was developed to describe the evolution of the full weight function of the color glass condensate. The BK equation \cite{Balitsky:1995ub,Kovchegov:1999yj,Kovchegov:1999ua} can be derived as a mean-field approximation of the JIMWLK hierarchy, formulated based on the unitarity of the dipole scattering amplitude.

	\subsection{Balitsky-Kovchegov equation}
	In this work, we employ the BK equation to characterize the saturation behavior of the proton and compare it with the criteria for a saturon system. In this work, we employ the BK equation to characterize the saturation behavior of the proton. Let $\mathcal{N}(k,Y)$ denote the dipole scattering amplitude in momentum space, which depends on the transverse momentum $k$ and the evolution rapidity $Y = \ln(1/x)$. Equation (\ref{eq:BK-int}) represents the momentum-space formulation of the BK equation, which is related to the coordinate-space version in Refs. \cite{Balitsky:1995ub,Kovchegov:1999yj} via a Fourier transform (see, e.g., Ref. \cite{Enberg:2005cb}). It is expressed as: \cite{Balitsky:1995ub,Kovchegov:1999yj}
	\begin{equation}
		\begin{aligned}
			\frac{\partial\mathcal{N}(k,Y)}{\partial Y}&=\bar{\alpha}_s\int_0^\infty\frac{\mathrm{d}\ell^2}{\ell^2}\Bigg[\frac{\ell^2\mathcal{N}(\ell,Y)-k^2\mathcal{N}(k,Y)}{|k^2-\ell^2|}\\
			&+\frac{k^2\mathcal{N}(k,Y)}{\sqrt{4\ell^4+k^4}}\Bigg]-\bar{\alpha}_s\mathcal{N}^2(k,Y).
		\end{aligned}
		\label{eq:BK-int}
	\end{equation}
	In this formulation, $N_c$ denotes the number of colors, $\alpha_s$ is the QCD coupling constant, and $\bar{\alpha}_s = \alpha_s N_c/\pi$. In addition, there exists a differential form of the BK equation, expressed using the BFKL kernel $\chi(\lambda)$, as given in Ref. \cite{Kovchegov:1999ua}
	\begin{equation}
		\begin{aligned}
			\frac{\partial\mathcal{N}(k,Y)}{\partial Y}&=\bar\alpha_s\chi\left(-\frac{\partial}{\partial\ln k^{2}}\right)\mathcal{N}(k,Y)\\&-\bar\alpha_s\mathcal{N}^{2}(k,Y).
		\end{aligned}
		\label{eq:BK-diff}
	\end{equation}
	The BFKL kernel in this context can be represented in terms of the digamma function $\psi(\lambda)=\Gamma^\prime(\lambda)/\Gamma(\lambda)$
	\begin{equation}
		\chi(\lambda)=\psi(1)-\frac{1}{2}\psi\left(1-\frac{\lambda}{2}\right)-\frac{1}{2}\psi\left(\frac{\lambda}{2}\right).
	\end{equation}
	
	While these two forms are mathematically equivalent representations of the same physical evolution, for our analysis we explicitly treat the dipole amplitude $\mathcal{N}(k,Y)$ as a function of the evolution rapidity $Y$ and the gluon transverse momentum $k$. Both formulations of the BK equation admit solutions in momentum space and can serve as inputs for calculating scattering amplitudes in diffractive processes as well as proton structure functions. Once the dipole amplitude is obtained from solving the BK equation, one can proceed to compute the unintegrated gluon distribution (UGD) and the gluon number density. Therefore, the solution to the BK equation captures the gluon evolution dynamics in the small-$x$ regime. For the nonlinear integro-differential form of Eq. (\ref{eq:BK-int}), a Runge-Kutta scheme incorporating integral terms can be employed. In this work, we use the BKsolver algorithm as outlined in Ref. \cite{Enberg:2005cb}, which approximates the integral term using Chebyshev polynomials. This method proves highly effective and allows for both fixed and running coupling scenarios. For the differential form of the BK equation, it is also possible to obtain approximate analytical solutions. By transforming the equation into the form of the Fisher-Kolmogorov-Petrovsky-Piskunov (FKPP) equation \cite{Munier:2003sj,Munier:2003vc,Munier:2004xu,Iancu:2004es,Enberg:2005uv,Munier:2014bba,Mueller:2018zwx}, one can apply techniques developed for nonlinear partial differential equations to perform the analysis.
	
	The solution to the BK equation, along with its characteristic quantity-the saturation scale $Q_s^2$-can be directly extracted from the solution itself. From the perspective of analytical analysis, we adopt the homogeneous balance method employed in Refs. \cite{Yang:2020jmt,Wang:2020stj,Wang:2022jwh,Cai:2023iza} to solve the BK (or FKPP) equation, which includes treatments for both fixed and running QCD coupling. Details of the solution methodology can be found in the relevant literature.
	In the fixed coupling (fc) scenario, an analytical solution to the BK equation in momentum space takes the form given as
	\begin{equation}
		\mathcal{N}_{fc}(L,Y)=\frac{A_0\mathrm{e}^{5A_0Y/3}}{\left[\mathrm{e}^{5A_0Y/6}+\mathrm{e}^{\left[-\theta+\sqrt{A_0/6A_2}(L-A_1Y)\right]}\right]^2}
		\label{eq:ana-sol-fc}
	\end{equation}
	with $L=\ln(k^2/\Lambda_\mathrm{QCD}^2)$, where $A_{0,1,2}$ and $\theta$ are fitting parameters. These parameters can be determined using gluon distribution data or proton structure function data \cite{Wang:2020stj,Wang:2022jwh}, in this work we use $A_0=33.3,\ A_1=-58.3,\ A_2=26.2$, and
	$\theta=-3.09$ \cite{Wang:2020stj}; however, as this is not the primary focus of the present study, we do not elaborate further.
	To extract the saturation scale $Q_s$, which corresponds to the transverse momentum at the saturation threshold, we directly utilize the analytical solution via the method described in Ref. \cite{Wang:2020stj}
	\begin{equation}
		Q_\mathrm{s}^2(Y)=\Lambda_\mathrm{QCD}^2\mathrm{e}^{(A_1+5\sqrt{A_0A_2/6})Y}.
		\label{eq:qs2fc}
	\end{equation}
	The associated parameters are consistent with those of the solution, and $\Lambda_\mathrm{QCD}$, representing the infrared cutoff, is taken to be 0.2 GeV.
	
	For the case of running coupling (rc), the details of the treatment using the homogeneous balance method can similarly be found in Ref. \cite{Cai:2023iza}. In this work, we adopt the saturation scale prescription (SSP). We note that while a rigorous derivation of the running coupling in small-$x$ evolution requires a loop computation to resum vacuum polarization bubbles as established by Balitsky \cite{Balitsky:2006wa}, the SSP serves here as a phenomenological ansatz whereby the running coupling scale is identified with the saturation scale $Q_s$. This choice reflects the characteristic momentum of gluons in the saturated regime and offers improved performance in describing the small-$x$ proton structure function $F_2$. The scale-dependent running coupling constant is expressed as \cite{Cai:2023iza}
	\begin{equation}
		\bar{\alpha}_s(Q_s^2)=\frac{1}{b\ln\frac{Q_s^2}{\Lambda_\mathrm{QCD}^2}},
	\end{equation}
	where $b=(11N_c-2N_f)/12N_c$. Note that this coefficient differs from the standard QCD $\beta$-function coefficient by a factor of $N_c/\pi$, arising from the definition of the reduced coupling $\bar{\alpha}_s = (N_c/\pi)\alpha_s$.
	
	The saturation scale $Q_s$ is proportional to the gluon density and thus exhibits a positive correlation with the rapidity $Y$. This relationship can be established through \cite{Iancu:2002tr}
	\begin{equation}
		Q_s^2(Y)=\Lambda_\mathrm{QCD}^2 \mathrm{e}^{\sqrt{cY}}
		\label{eq:qs2rc}
	\end{equation}
	with the free parameter $c$. The transition from the exponential growth $e^{\lambda Y}$ in the fixed coupling case to the slower $e^{\sqrt{Y}}$ behavior in the running coupling scenario has a clear physical origin. As the saturation scale $Q_s(Y)$ increases with rapidity, the effective coupling $\bar{\alpha}_s(Q_s^2)$ decreases. This dynamical reduction in the coupling strength suppresses the gluon emission probability, thereby slowing down the evolution speed of the saturation wavefront compared to the fixed coupling case. This is consistent with the reasoning behind Eq. (\ref{eq:qs2fc}), where, upon assuming a specific scale for the coupling constant, the analytical solution to the BK equation in the running coupling scenario can also be obtained using the homogeneous balance method. The explicit form is given as \cite{Cai:2023iza}
	\begin{widetext}
	\begin{equation}
		\mathcal{N}_{rc}(L,Y)=\frac{A_0\exp\left[2\theta+2\frac{\left(\frac{\sqrt{6A_0}A_1}{\sqrt{A_2}}+5A_0\right)\sqrt{cY}}{3bc}\right]}{\left\{\exp\left[\theta+\frac{\left(\frac{\sqrt{6A_0}A_1}{\sqrt{A_2}}+5A_0\right)\sqrt{cY}}{\sqrt{A_2}}\right]+\exp\left[\frac{\sqrt{A_0}L}{\sqrt{6A_2}}\right]\right\}^2},
		\label{eq:ana-sol-rc}
	\end{equation}
    \end{widetext}
	Similarly, this analytical solution contains free parameters that must be determined by fitting to experimental data. Specifically, in the case of running coupling, the parameters of the analytical solution are obtained by fitting to the experimental data of the structure function \cite{H1:2009pze,H1:2013ktq}, yielding $A_0=6.98,\ A_1=-8.46,\ A_2=2.95,\ c=3.11$, and $\theta=-2.76$.
	
	A detailed discussion of the numerical solutions to the BK equation is beyond the scope of this work; here we briefly outline the method for extracting the saturation scale $Q_s^2$ from such solutions. While the standard definition of the saturation scale is typically formulated in coordinate space (as the inverse radius $r=1/Q_s$ where the amplitude reaches a fixed value), it is operationally natural in our momentum-space analysis to extract $Q_s$ directly from the momentum-dependent amplitude. We verify that for appropriate threshold choices (e.g., $\kappa \sim 0.5$), this momentum-space extraction reproduces the correct universal asymptotic behavior at large rapidity, consistent with theoretical expectations derived in coordinate space \cite{Munier:2003vc}. Notably, for $\kappa=1$, the numerical solutions obtained using BKsolver \cite{Enberg:2005cb} reproduce the expected asymptotic behavior well in both fixed and running coupling scenarios.
	
\begin{figure*}[htpb]
		\begin{center}
			\includegraphics[width=0.45\textwidth]{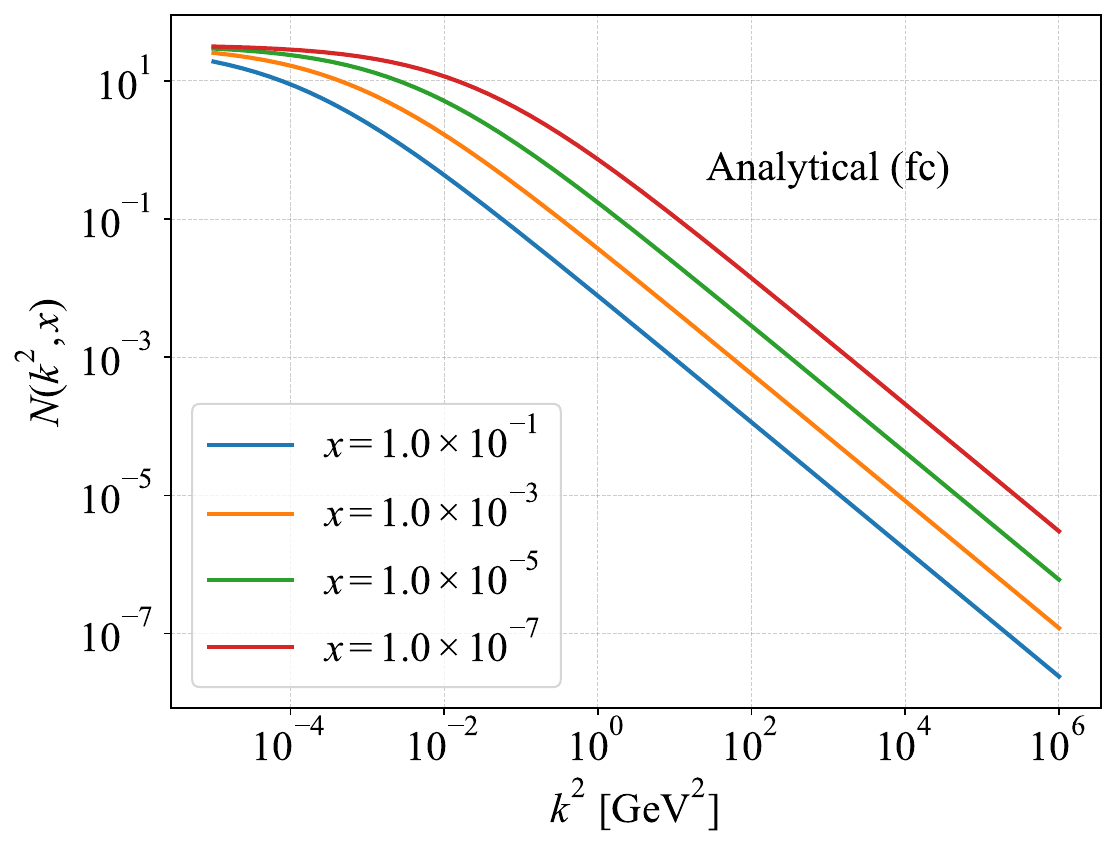}
			\includegraphics[width=0.45\textwidth]{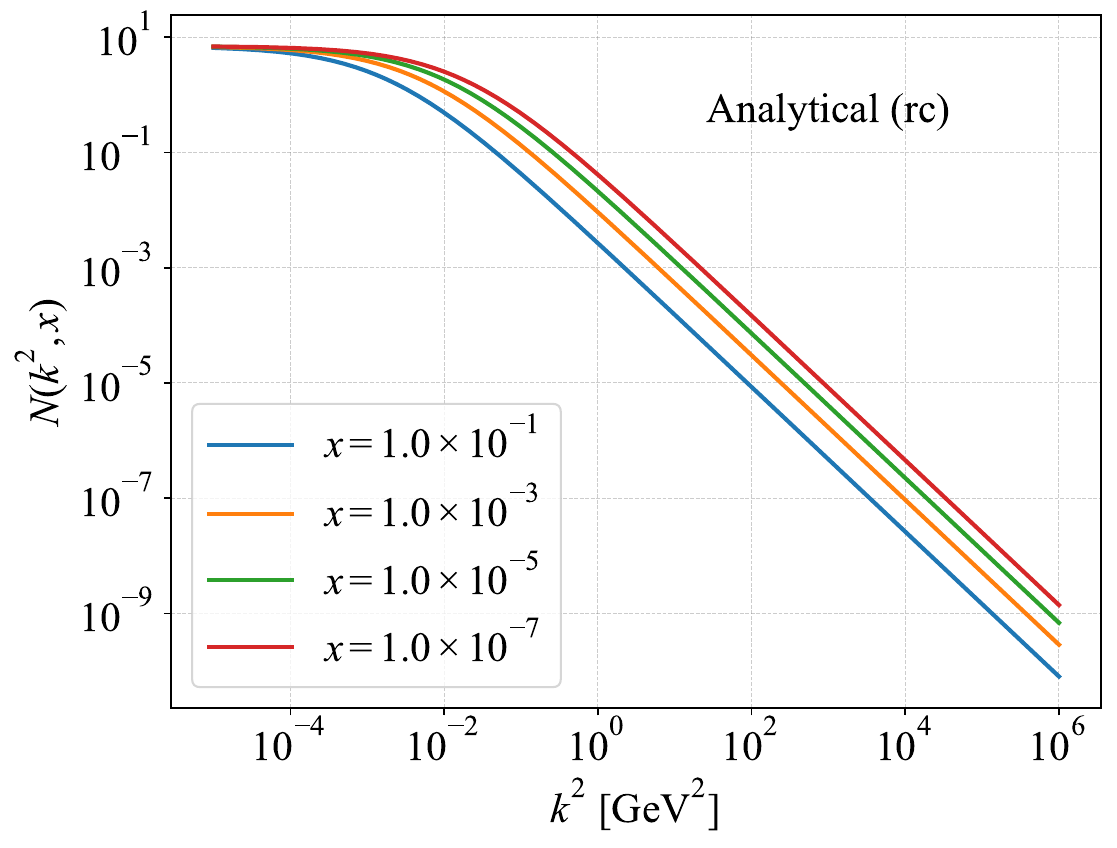}
			\includegraphics[width=0.45\textwidth]{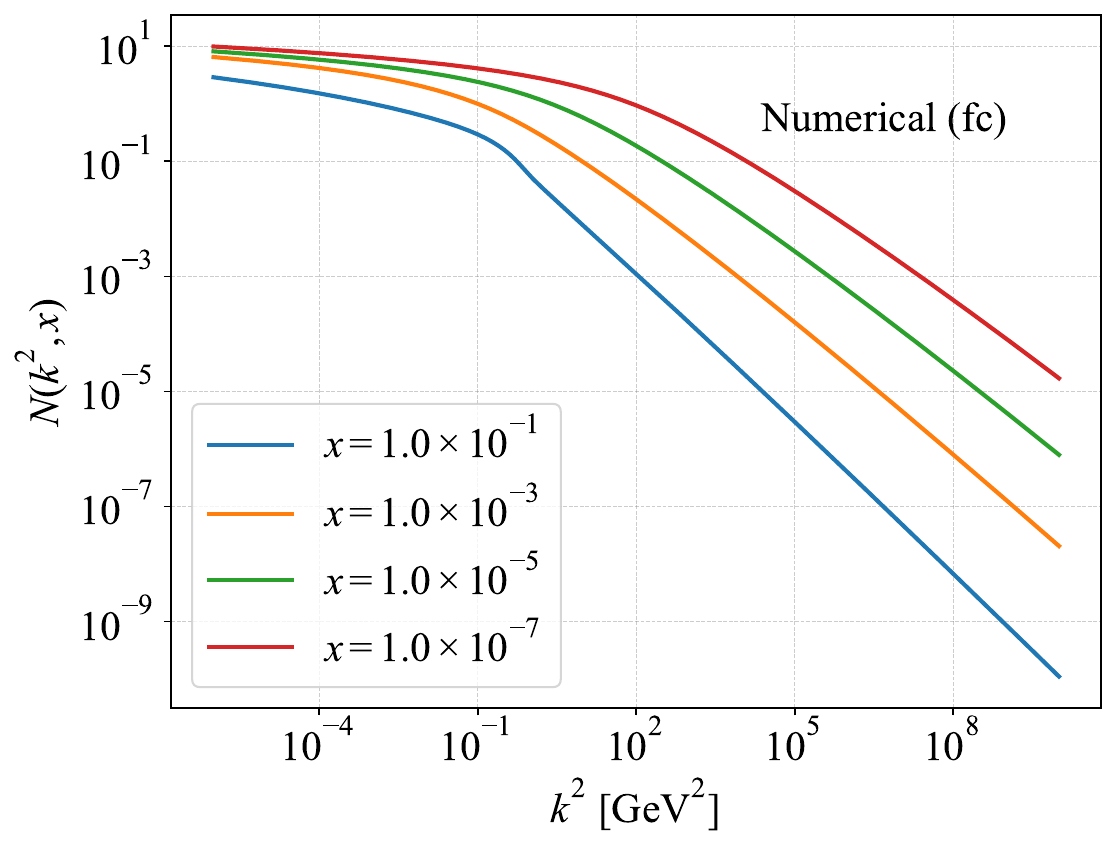}
			\includegraphics[width=0.45\textwidth]{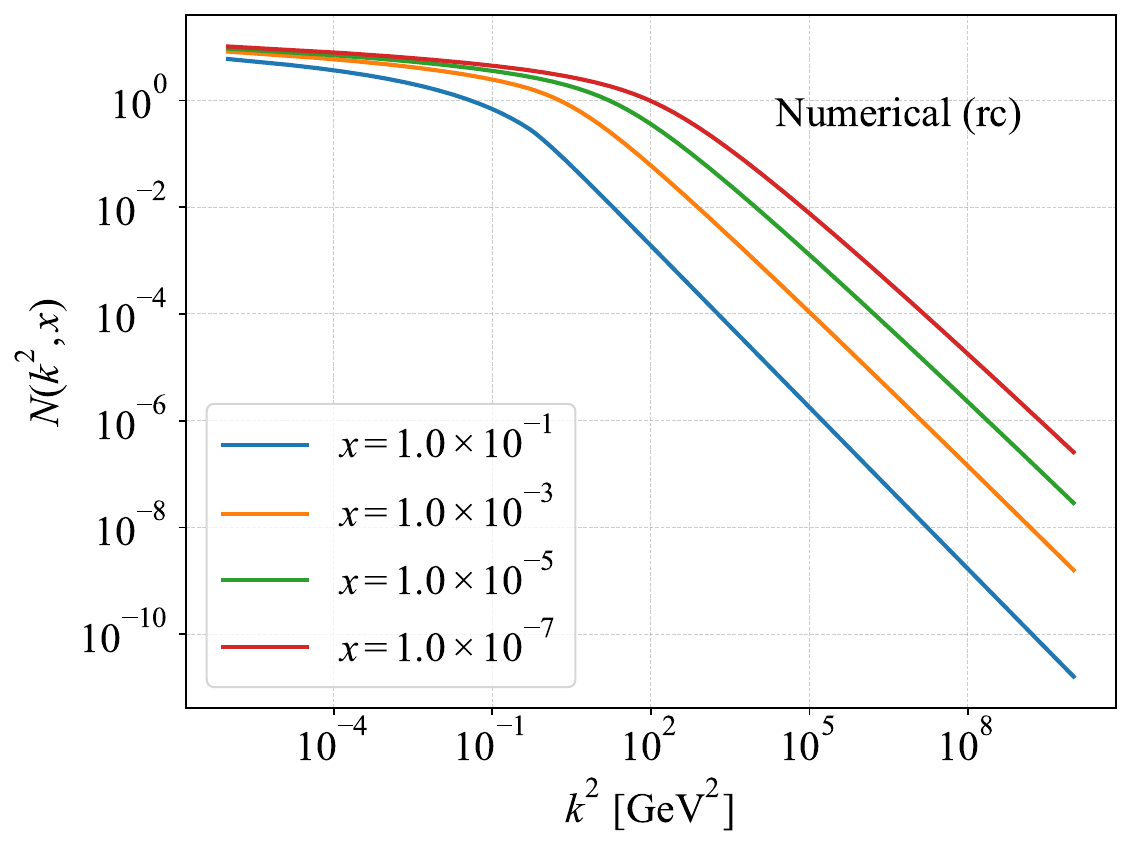}
		\end{center}
		\caption{From left to right and top to bottom, the panels correspond to the analytical and numerical solutions of the BK equation in the fixed and running coupling cases, respectively. The solution methods follow \cite{Wang:2020stj,Cai:2023iza,Enberg:2005cb}. In each subfigure, the four curves from left to right represent $x = 10^{-1},\ 10^{-3},\ 10^{-5},\ 10^{-7}$, respectively.}
		\label{fig:sol}
	\end{figure*}

\begin{figure}[htpb]
		\begin{center}
			\includegraphics[width=0.45\textwidth]{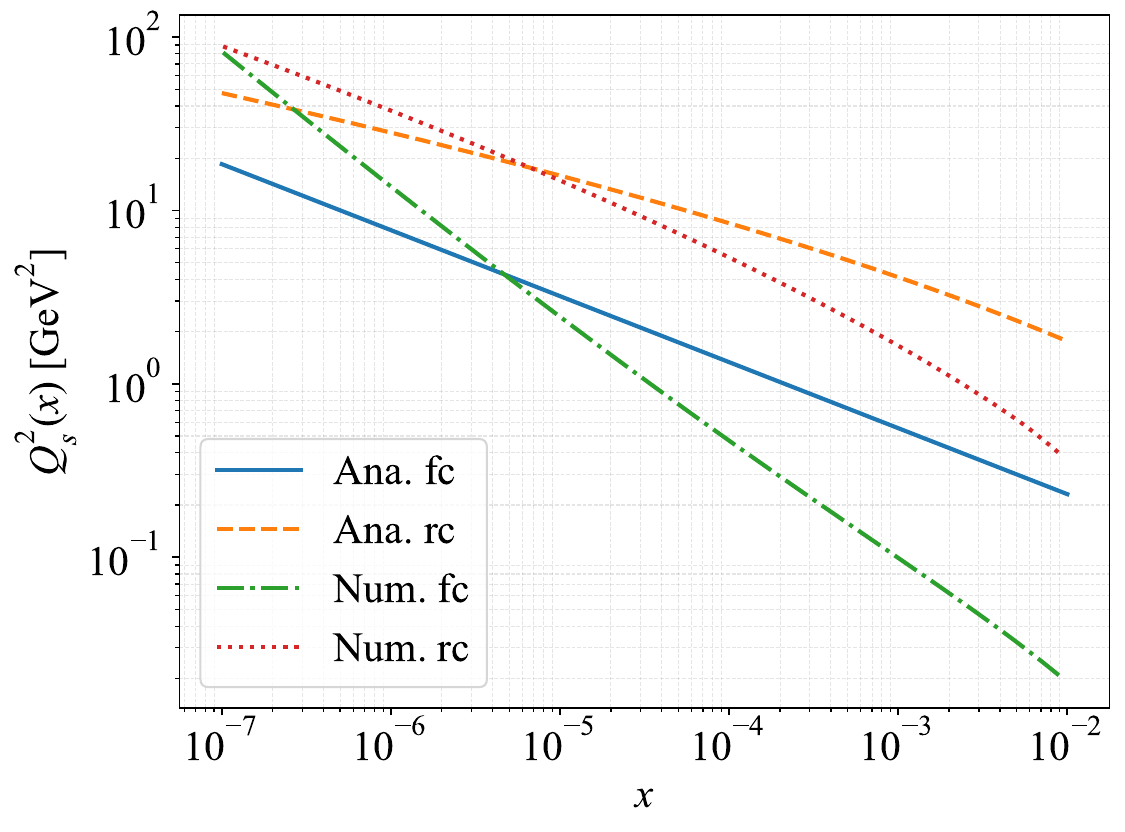}
		\end{center}
		\caption{The variation of the extracted saturation scale $Q_s^2$ with respect to $x$ under different schemes is shown. The solid line, dashed line, dash-dotted line, and dotted line correspond to the four respective solutions.}
		\label{fig:qs2}
	\end{figure}

	Figure \ref{fig:sol} presents both the analytical and numerical solutions of the BK equation with fixed and running coupling, obtained using the methods described above. The figure displays the distribution of the dipole amplitude as a function of gluon transverse momentum for various values of $x$ (In each subfigure, the four curves from left to right correspond to $x = 10^{-1},\ 10^{-3},\ 10^{-5},\ 10^{-7}$, respectively.
	). It is evident that all solutions exhibit wavefront-like profiles, a feature connected to the statistical properties underlying the BK equation. The analytical solutions, in particular, provide explicit functional forms for these traveling wave solutions.
	Furthermore, FIG. \ref{fig:qs2} illustrates the evolution of the saturation scale $Q_s^2$ as a function of $x$ for different solution schemes. The solid line, dashed line, dash-dotted line, and dotted line correspond to the four respective solutions. While some discrepancies exist between the extracted saturation scales across methods, due to differing assumptions in solving the equation or defining the saturation condition, the general trend remains consistent: $Q_s^2$ increases as $x$ decreases.

	\subsection{Saturons, occupation number and entropy}
	As discussed earlier, we aim to determine whether high-occupancy systems exhibit distinctive dynamical properties at the critical point. Taking black holes as an example, critical packing leads to the emergence of a large number of gapless modes, which correspond to specific microstate entropy $S$. Among the relations involving entropy, the most significant is its dependence on the system coupling constant $\alpha(Q_s)$ with saturation scale $Q_s$ at the critical point. This correspondence was proposed in Refs. \cite{Dvali:2019jjw,Dvali:2019ulr,Dvali:2020wqi} and implies an upper bound on the microstate entropy and occupation number of a system
	\begin{equation}
		S_{\max} \simeq N_{\max}=\frac{1}{\alpha(Q_S)}.
		\label{eq:entropy-bound}
	\end{equation}
	At the same time, Ref. \cite{Dvali:2020wqi} further proposed that the correlation between unitarity-preserving saturation and the entropy bound (\ref{eq:entropy-bound}) is inherently nonperturbative and cannot be circumvented through resummation schemes within perturbative frameworks. Moreover, any system that reaches this entropy limit exhibits, to some extent, universal features shared with black holes. In the relevant literature, the connection between black holes and analogous saturated objects in other physical systems has been established, including states in gauge theories with a large number of colors. 
	
	Several studies have suggested a correspondence between black holes and the CGC formed in high-energy matter collisions, the latter being a high-occupancy condensate permitted within the framework of QCD. Although the critical behavior of QCD differs from that of gravity, they share common features in terms of critical packing. Both the black hole state and the CGC state emerge as outcomes of the classicalization and unitarization of the $2 \to N$ amplitude in the high-energy Regge limit of gravity or QCD \cite{Dvali:2021ooc}. The interpretation of black holes and other field-theoretic systems as saturated states has been extensively discussed in the literature. Our focus here is on the nucleon system, more precisely the high-energy proton parton distribution system governed by QCD. We investigate whether it can be regarded as a saturated state, and examine its implications for microscopic entropy as well as the definition of high-occupancy states.
	
	One may consider a highly occupied proton, namely the gluon distribution inside the proton in the small-$x$ region. This distribution can be described within the framework of the BK equation through the relation between the dipole scattering amplitude and the UGD. The gluon occupancy $N_g$ at high energies can be expressed in terms of the gluon multiplicity distribution produced in collision processes, given as \cite{Gelis:2010nm,Iancu:2003xm}
	\begin{equation}
		N_g(x,\boldsymbol{k}_\perp,\boldsymbol{x}_\perp)\equiv\frac{(2\pi)^3}{2(N_c^2-1)}\mathrm{~}x\frac{\mathrm{d}N_\mathrm{gluon}}{\mathrm{d}x\mathrm{~d}^2\boldsymbol{k}_\perp\mathrm{d}^2\boldsymbol{x}_\perp},
		\label{eq:occupy-number1}
	\end{equation}
	where $\boldsymbol{k}_\perp$ and $\boldsymbol{x}_\perp$ are transverse momentum and impact parameter of gluon, respectively. Eq. (\ref{eq:occupy-number1}) defines the microscopic gluon occupation number in phase space. The standard collinear gluon distribution $xG(x,Q^2)$ is obtained by integrating this density:
\begin{equation}
    xG(x,Q^2) = \int^{Q^2} \frac{d^2k_\perp}{(2\pi)^2} \int d^2b_\perp \, 2(N_c^2-1) N_g(x,k_\perp,b_\perp).
\end{equation}
Assuming a homogeneous transverse density profile within a proton of radius $R$, the integration over impact parameter yields a geometric factor of $\pi R^2$. We can thus invert this relation to define an average gluon occupation number:
\begin{equation}
    N_g(x,Q^2) \simeq \frac{xG(x,Q^2)}{\pi R^2 Q^2}.
    \label{eq:occupy-number2}
\end{equation}
Regarding the unintegrated gluon distribution (UGD) $\mathcal{F}(x,k^2)$, we adopt a form inspired by Ref. \cite{Kutak:2004ym}:
\begin{equation}
    \mathcal{F}(x,k^2) = \frac{N_c R^2}{\alpha_s \pi} \left(1-k^2\frac{d}{dk^2}\right)^2 k^2 \mathcal{N}(k,Y).
    \label{eq:UGD}
\end{equation}
We note that while Ref. \cite{Kutak:2004ym} utilizes an auxiliary function without the explicit area factor, we include the term $\pi R^2$ here to ensure dimensional consistency with the definition of the average occupation number in Eq. (\ref{eq:occupy-number2}).
	
	We now turn to the discussion of how to define the concept of saturation entropy for the proton in a high gluon occupancy regime. A potential approach to resolving the entropy of such a highly occupied system was proposed in  \cite{Kharzeev:2005iz,Castorina:2007eb} and subsequently applied in \cite{Kutak:2011rb} to address entropy production in high-energy collisions. This perspective originates from the partonic picture of QCD, where the thermalization of a hadron is associated with a temperature linked to its saturation scale
	\begin{equation}
		T=\frac{Q_s(x)}{2\pi},
		\label{eq:unruh}
	\end{equation}
	where $x$ stands for a longitudinal momentum fraction of hadron
	carried by a parton. This connection is established through the well-known Unruh effect \cite{Unruh:1976db}. A concise formulation of the Unruh effect states that an observer undergoing constant acceleration $a$ perceives a thermal bath with temperature $T = a / 2\pi$ in their own rest frame. When applied to off-shell gluons being radiated, this suggests a deceleration effect in the transverse-momentum-dependent UGD. By analyzing gluon deceleration in a uniform chromoelectric field, one finds that the characteristic acceleration is compatible in magnitude with the saturation scale \cite{Kharzeev:2005iz,Wong:1970fu}, such that $a = Q_s(x)$
	
	In addition to temperature, a complete thermodynamic definition of entropy requires a description of energy. Inspired by \cite{Kovchegov:1999yj,Kovchegov:1999ua}, we associate the emergent mass generated through the nonlinear evolution of gluons with the saturation scale. Specifically, this mass is taken to be proportional to the saturation scale, thereby avoiding divergences in the gluon propagator. Furthermore, it has been suggested that gauge-invariant gluons may acquire a dynamical mass due to nonlinear effects (see Refs. \cite{Mathieu:2010nbc,Mathieu:2011zzb} and references therein). Accordingly, one may express the mass of a single gluon as \cite{Kutak:2011rb}
	\begin{equation}
		M_g(x) = Q_s(x).
		\label{eq:gluon-mass}
	\end{equation}
	
	Based on the definitions of gluon mass and system temperature discussed above, one can formulate a coarse-grained entropy, as implied by the fundamental laws of thermodynamics, under the assumption that the system volume remains approximately constant. Using the energy-or the effective mass of the gluons-together with the temperature derived from the Unruh effect, the entropy is found to satisfy the following differential equation
	\begin{equation}
		dM=dE=TdS,
		\label{eq:entropy}
	\end{equation}
	here, the mass $M$ can be interpreted as the total emergent mass of all gluons, namely, the total energy of $N_g$ gluons with individual mass $M_g(x)$, such that $M(x) = N_g(x) M_g(x)$. Naturally, the longitudinal momentum fraction $x$ carried by the gluons should be accounted for in the dependence of the above quantities, and it can serve as one of the scaling variables for the entropy. By introducing the Unruh acceleration $a = Q_s(x)$, the differential expression for the entropy can then be obtained directly from Eq. (\ref{eq:unruh}) as follows:
	\begin{equation}
		dS=2 \pi \frac{d\left[N_g(x)M_g(x)\right]}{Q_s(x)}.
		\label{eq:final-entropy}
	\end{equation}
	For simplicity, the lowest entropy state $S_0$ can be set to zero. All integration constants can be absorbed into quantities related to the gluon occupation number, which may be determined from experimental data or global QCD fits. It is important to emphasize that, in this analysis, the gluon occupation number (or gluon density) is obtained from the solution of the BK equation. Ultimately, following the thermodynamic definition proposed by Kutak in Ref. \cite{Kutak:2011rb}, the entropy of the system takes the form:
	\begin{equation}
		S(x)\simeq\pi N_g(x) \ln\frac{Q_s^2(x)}{\Lambda_\mathrm{QCD}^2}.
		\label{eq:final-entropy2}
	\end{equation}
	The thermodynamic entropy of the high-occupancy gluon state obtained above is consistent with the conclusion presented in \cite{Kutak:2023cwg}, where a simplified analytic expression for the gluon distribution based on the Golec-Biernat Wuesthoff (GBW) model \cite{Golec-Biernat:1998zce} yields a result that matches the entanglement entropy derived from DIS \cite{Kharzeev:2017qzs,Hentschinski:2024gaa,Kou:2022dkw,Hentschinski:2021aux,Hentschinski:2022rsa,Hentschinski:2023izh,H1:2020zpd,Tu:2019ouv}.
	
	The properties of black holes as saturons systems have been extensively studied. We now turn to the question of whether a highly saturated proton, characterized by a large gluon occupation number in its interior, can exhibit features analogous to those of black holes. As previously discussed, saturons systems display certain universal behaviors near the critical regime. The Bekenstein-Hawking entropy of a black hole suggests that for a black hole of radius $R$, the entropy scales as $S \sim \mathrm{Aera}/G_{gr}$, where $G_{gr}$ denotes Newton’s gravitational constant and has dimensions of $R^{d-2}$. The saturation property of black holes leads to the entropy saturating the Bekenstein-Hawking bound, yielding $S \sim 1/\alpha_{gr}$ with $\alpha_{gr} = G_{gr}/R^{d-2}$.
	
	We now consider the internal dynamics of a highly saturated proton governed by QCD. The gluon distribution at the saturation scale $Q = Q_s(x)$, denoted by $xG(x, Q_s(x))$, can be obtained by integrating the UGD over transverse momentum. The UGD itself is derived from the solution of the BK equation via the dipole scattering amplitude and Eq. (\ref{eq:UGD}). It can be shown that the gluon distribution obtained from the UGD satisfies $xG \sim A_\perp / \alpha_s$ \cite{Golec-Biernat:1998zce,Kutak:2023cwg}, where $A_\perp = \pi R^2$ represents the transverse area of the boosted proton. The gluon occupation number $N_g$ is related to $xG$ through Eq. (\ref{eq:occupy-number2}) and ultimately enters the expression for the system's thermodynamic entropy via Eq .(\ref{eq:final-entropy2}), resulting in the scaling 
	\begin{equation}
		S(x) \sim N_g(x) \sim 1/\alpha_s.
		\label{eq:final-compare}
	\end{equation}
	This provides the basis for considering the high-energy, high-occupancy gluon state in the proton as a saturons system. In contrast to the gravitational case, the relevant coupling constant is now the QCD strong coupling $\alpha_s$, which depends on the saturation scale as $\alpha_s \sim \alpha_s(Q_s^2)$.
	
	\subsection{From proton to nucleus: Glauber-Gribov-Mueller Scheme}
	It is also worth mentioning the extension of the nucleon UGD to the nuclear case, since the nuclear mass number ($A$) can enhance the gluon distribution in the high-saturation regime of a nucleus, in line with the CGC picture. In this work we adopt a simple scheme that maps the UGD from a single proton to a high-mass-number nucleus (e.g., Pb), with the goal of straightforwardly comparing its saturation criterion with that of a single nucleon. Multiple scattering in a dense nuclear target can be described by the Glauber-Gribov-Mueller (GGM) approximation \cite{Glauber:1955qq,Gribov:1968gs,Gribov:1968jf,Mueller:1989st,Mueller:1999wm,Kovchegov:2014kua}. The nuclear saturation scale grows with mass number \cite{Gelis:2010nm,Kovchegov:2014kua}
	\begin{equation}
		Q_{s,A}^2(x)\;\simeq\;\,A^{1/3}\,Q_{s,p}^2(x),\qquad
		R_A=r_0\,A^{1/3},
		\label{eq:QsA_scaling}
	\end{equation}
	where $Q_{s,p}=Q_s$ in Eq. (\ref{eq:qs2fc}) represents the nucleon saturation scale, $R_A$ is the nuclei radius with mass number $A$ and $r_0=1.2$ fm. We note that the scaling $Q_{s,A}^2 \sim A^{1/3}$ assumes a simplified cylindrical geometry corresponding to the maximum optical depth in central collisions. As pointed out in recent phenomenological studies \cite{Lappi:2013zma,Deganutti:2023qct}, a more realistic treatment involving the Woods-Saxon nuclear density profile and averaging over impact parameters (particularly for minimum-bias events) leads to a significantly more modest growth of the saturation scale. Therefore, the estimate in Eq. (\ref{eq:QsA_scaling}) should be regarded as an upper bound on the nuclear enhancement. However, even with a softer scaling, the saturation scale of a heavy nucleus remains substantially larger than that of a proton, preserving the qualitative ordering discussed in this work.
	
	Unlike the dipole scattering amplitude for the proton -- i.e., the solution of the BK equation -- the nuclear dipole amplitude is taken in the eikonalized GGM form \cite{Gribov:1968gs,Gribov:1968jf,Armesto:2003pq}
	\begin{equation}
		N_A(r,b,x)=1-\exp\Big[-\tfrac{1}{2}\sigma_{\rm dip}^p(r,x)T_A(b)\Big].
		\label{eq:GM_nucleus}
	\end{equation}
	Here, $T_A$ denotes the Woods-Saxon nuclear thickness profile \cite{Woods:1954zz}, which together with the dipole-proton scattering cross section $\sigma_{\rm dip}^p$ composes the dipole-nucleus scattering amplitude.
	
	The $\sigma_{\rm dip}^p$ is obtained from a Hankel transform of the proton UGD, which maps momentum-space information into coordinate space. The proton UGD has the form given above, explicitly written as \cite{Kovchegov:2014kua,Kovchegov:2012mbw}
	\begin{equation}
		\begin{aligned}
			\sigma_{\mathrm{dip}}^p(x,r)&=\frac{8\pi\alpha_s}{N_cR_p^2}\int_0^\infty dk\frac{1-J_0(kr)}{k^3}\mathcal{F}_p(x,k^2)
		\end{aligned}
	\end{equation}
	Note that $N_A$ is the nuclear amplitude in the fundamental representation. To move to the adjoint representation while enforcing unitarity \cite{Kutak:2011rb}, deform it as
	$N_G(r,b,x)=2N_A(r,b,x)-N_A^2(r,b,x)$.
	Then integrate over the impact parameter $b$; applying the gradient and performing a Fourier transform yields the UGD for the nuclear case
	\begin{equation}
		\mathcal{F}_A(x,k^2)=\frac{N_cR_A^2\pi}{2\alpha_s^2C_F}k^2\phi_A(k,x)
	\end{equation}
	with
	\begin{equation}
		\begin{aligned}
			\phi_A(k,x)&=\frac{C_F\alpha_s}{(2\pi)^3}\int d^2re^{-ik\cdot r}\nabla_r^2\left[\int d^2bN_G(r,b,x)\right].
		\end{aligned}
	\end{equation}
	With the nuclear UGD in hand, one can, by analogy with the proton case, compute the nuclear gluon occupancy, thermodynamic entropy, and related quantities. This enables a comparison with the saturation criterion. We will present a more detailed discussion in the next section.

	\section{results and discussions}
	\label{sec:diss}
	Up to this point, we have formally discussed the potential of high-energy protons to serve as candidates for saturons. However, accurately determining the properties of saturons from gluon distributions remains a substantial challenge, owing to higher-order corrections in the small-$x$ gluon evolution within the proton and certain non-perturbative effects. Additionally, the number of gluons radiated by a single proton is considerably smaller than that found in the highly saturated color-charge states produced in heavy-ion collisions. Nevertheless, it is of significant interest to assess whether protons can be identified as saturons through their gluon distributions. A suitable framework for describing saturated protons is provided by the small-$x$ QCD evolution equations, specifically the nonlinear BK/JIMWLK equations. These equations can be expressed in terms of the gluon occupation number, with the explicit nonlinear form given by $\partial_Y N_g =\omega\alpha_s N_g-\alpha_s N_g^2$, where $\omega$ is a number of order unity \cite{Gelis:2010nm}. The nonlinear term originates from the gluon recombination process $gg \to g$. Upon reaching saturation, the gluon occupation number should become independent of the rapidity evolution variable $Y$, leading to $N_g \sim 1/\alpha_s$. This provides a criterion for identifying saturon systems directly from the evlution equation itself, namely Eq. (\ref{eq:final-compare}).

	\subsection{Proton case}
	Through the mean-field approximation of the nonlinear evolution equation, namely the momentum-space BK equation, both its analytical and numerical solutions for the dipole amplitude can be related to the gluon distribution via the definition of the UGD. This connection allows the determination of the gluon occupation number and the corresponding thermodynamic entropy, as given in Eqs (\ref{eq:occupy-number2}-\ref{eq:final-entropy2}). 
	
\begin{figure*}[htpb]
		\begin{center}
			\includegraphics[width=0.45\textwidth]{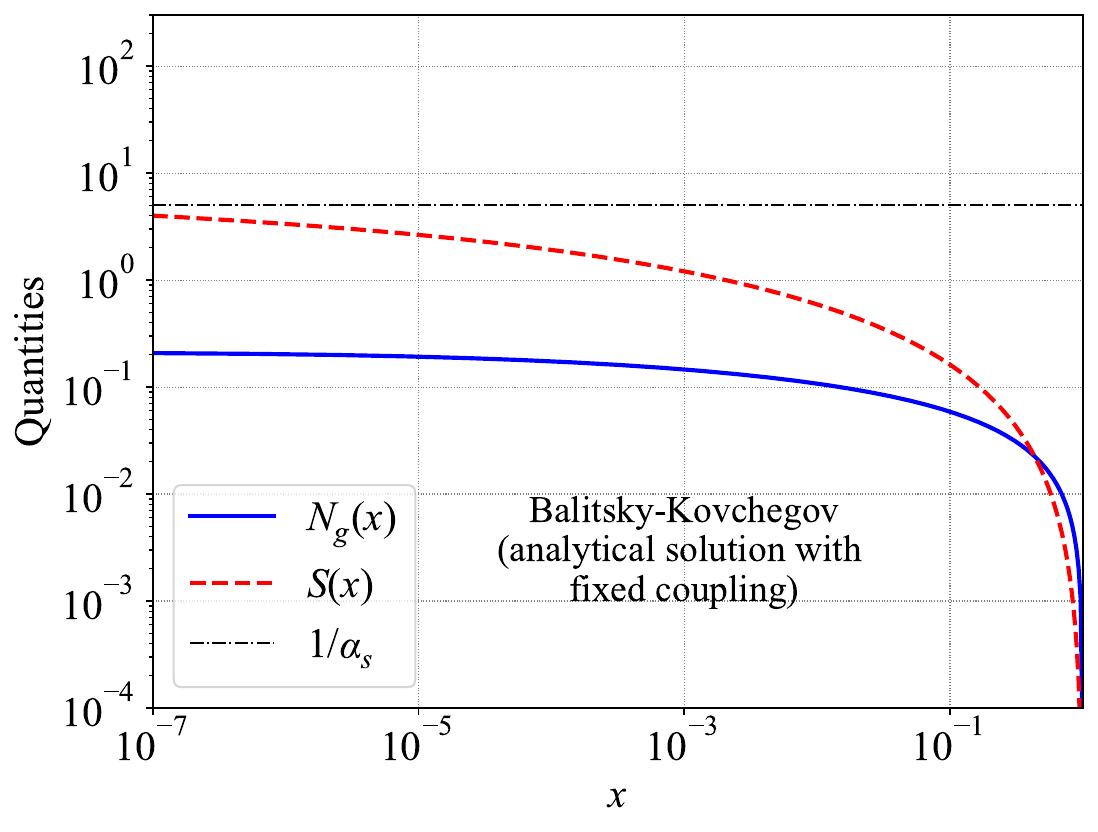}
			\includegraphics[width=0.45\textwidth]{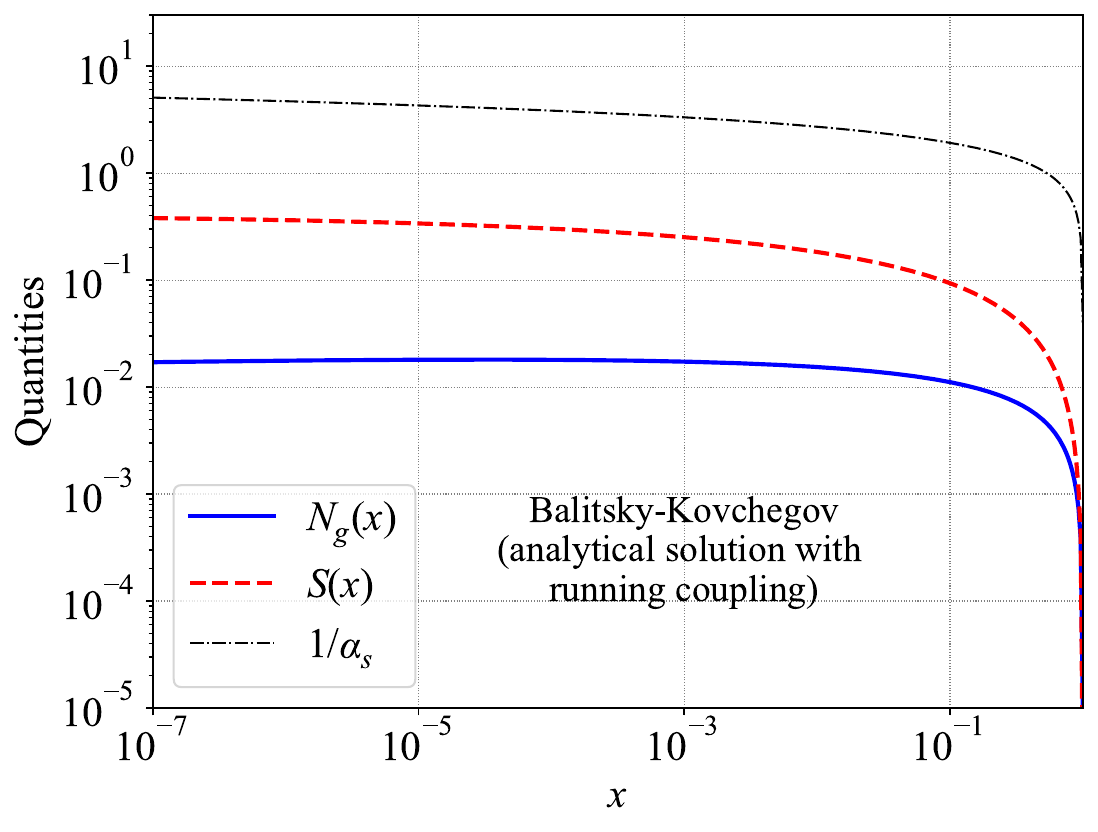}
			\includegraphics[width=0.45\textwidth]{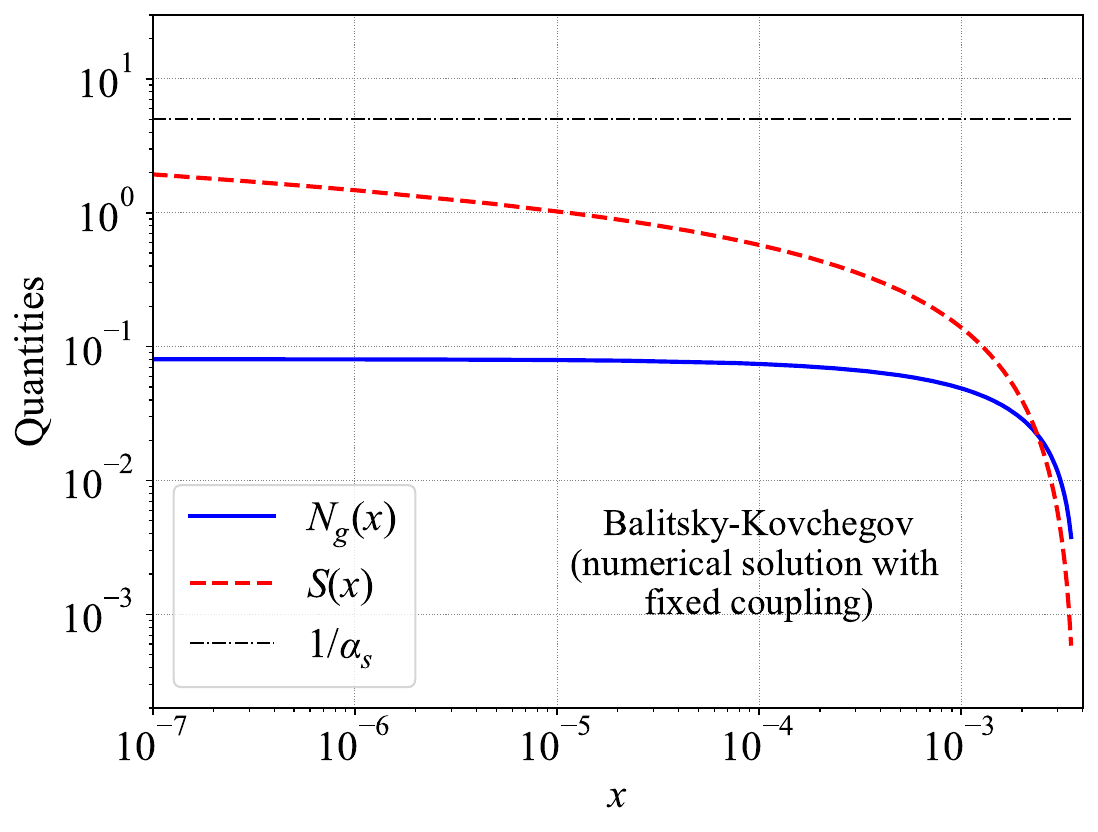}
			\includegraphics[width=0.45\textwidth]{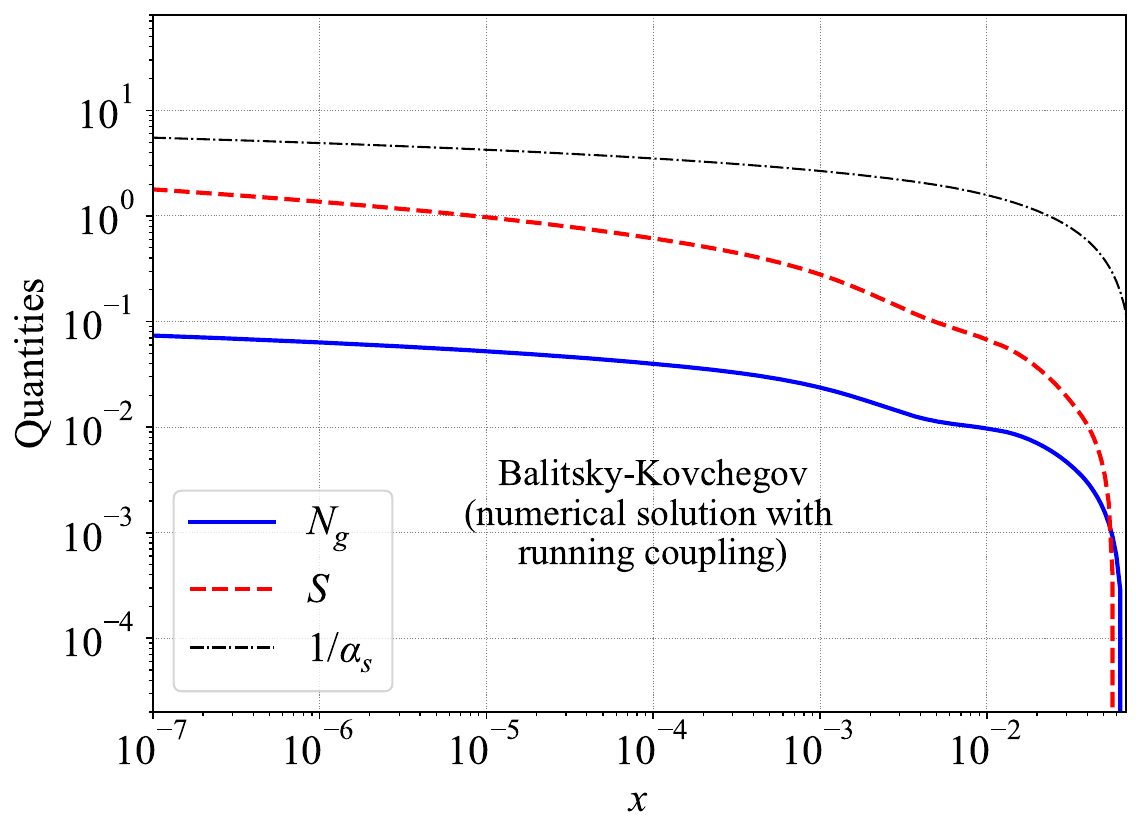}
		\end{center}
		\caption{The four subplots, arranged from left to right and top to bottom, represent respectively: the analytical solution of the BK equation with fixed coupling, the analytical solution with running coupling, the numerical solution with fixed coupling, and the numerical solution with running coupling. Each panel displays the resulting gluon occupation number and the associated entropy.}
		\label{fig:results}
	\end{figure*}

	Following the above framework, the numerical results are presented in Fig. \ref{fig:results}. This includes both analytical and numerical solutions obtained with fixed and running coupling, showing the distributions of the gluon occupation number (blue solid lines) and the corresponding thermodynamic entropy (red dashed lines) as functions of $x$. For comparison with the coupling strength, we also include the coupling constant curve as a black dotted-dashed line. The fixed coupling is set to $\alpha_s = 0.2$, while the running coupling is determined using the one-loop $\beta$-function.

	The parameter choices for all results shown in the figure are specified in the main text and are taken from the corresponding references. The four subplots, arranged from left to right and top to bottom, represent respectively: the analytical solution of the BK equation with fixed coupling, the analytical solution with running coupling, the numerical solution with fixed coupling, and the numerical solution with running coupling. Each panel displays the resulting gluon occupation number and the associated entropy. It is observed that in all cases, even down to $x = 10^{-7}$---well below the kinematic reach of current DIS experiments---neither the entropy nor the gluon occupation number approaches the inverse coupling constant, $1/\alpha_s$. We note that while the figures include trajectories extending to $x \sim 10^{-1}$ to visualize the initialization and continuity of the evolution, our physical conclusions regarding the saturon bound are strictly based on the behavior in the deep small-$x$ region ($x < 10^{-3}$) where the BK formalism is theoretically valid. One important reason for this is the presence of approximations in the construction of the entropy, and the omission of higher-order corrections in the evolution equations. 
    
    Regarding the sensitivity to initial conditions, it is well known that the absolute value of the saturation scale $Q_s$ depends on the parameters of the initial distribution at $Y=0$. However, a key feature of the BK evolution is the emergence of traveling wave solutions, where the asymptotic velocity of the wavefront is determined by the evolution kernel rather than the initial inputs. This implies that while varying the initial conditions might horizontally shift the saturation scale (and consequently the entropy curves), the rate of growth with rapidity is a robust prediction of the theory. Therefore, our conclusion that the growth rate is insufficient to reach the saturon limit within the accessible kinematic window remains valid across reasonable variations of the initial parameters.
	
	\subsection{Nuclear case}
\begin{figure}[htpb]
		\begin{center}
			\includegraphics[width=0.45\textwidth]{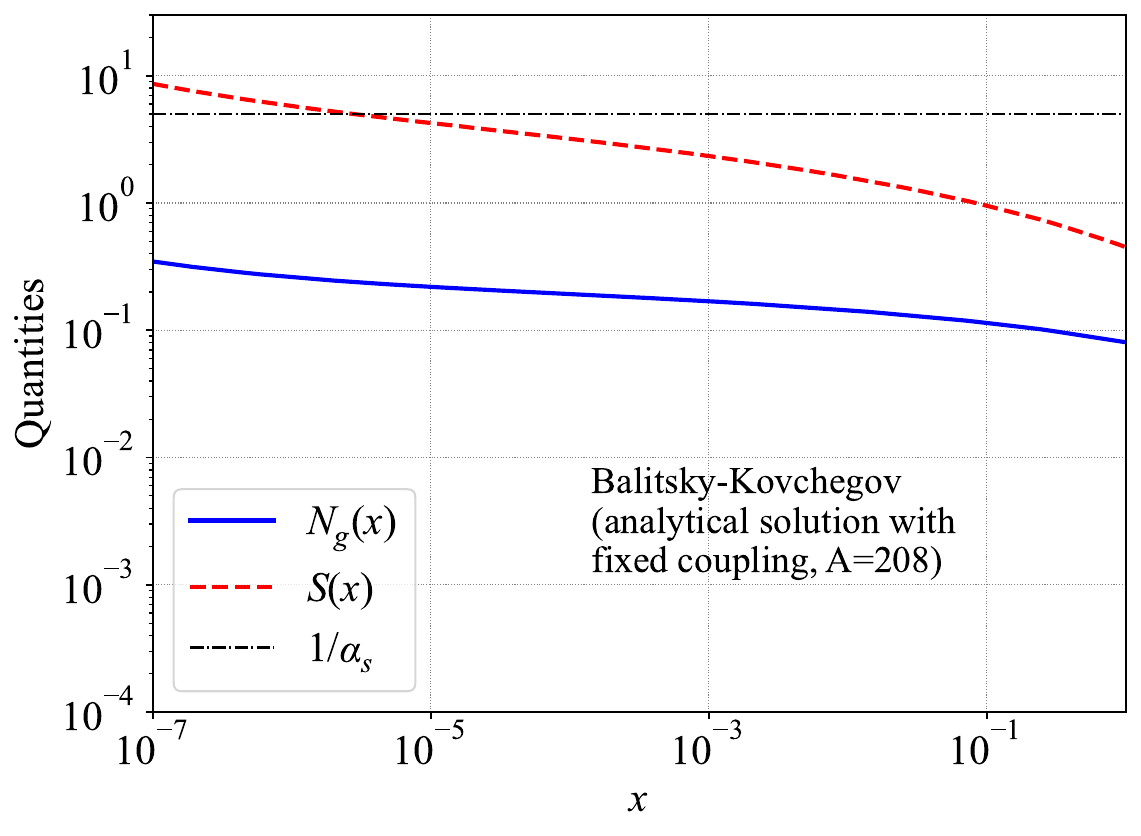}
		\end{center}
		\caption{Gluon occupancy $N_g(x)$ and thermodynamic entropy $S(x)$ for the nucleus ($A=208$) at fixed coupling; the black horizontal line denotes the criterion ($1/\alpha_s$).
		}
		\label{fig:nuc_results_fixed}
	\end{figure}

	A simple estimate is needed to assess how much a nucleus can enhance quantities such as entropy. We give a single example: the proton dipole amplitude is taken from an analytic solution of the BK equation and extended to the nuclear case using the GGM scheme. We choose a mass number $A=208$, corresponding to a lead nucleus.
	Using the same numerical procedure and parameter settings as for the proton (changing only the target from proton to nucleus ($A=208$) and fixing the coupling at $\alpha_s=0.2$), we obtain the nuclear target’s gluon occupancy $N_g(x)$ and thermodynamic entropy $S(x)$. Figure~\ref{fig:nuc_results_fixed} displays these two curves, with the reference line $1/\alpha_s$ overlaid (the same criterion as in the proton case).
	As seen in the figure, compared with the proton result, under the $x$-dependent growth of the entropy, the thermodynamic entropy of a highly occupied lead nucleus reaches the saturation criterion ($1/\alpha_s$) at around $x \sim 10^{-6}$. This indicates that the nucleus indeed enhances the entropy, mainly through the increase of the transverse radius after the high-energy boost (set by the nuclear radius) and the amplification of the nuclear saturation scale. However, this comparison has limitations, such as the choice of the dipole-amplitude form and uncertainties from nuclear geometric parameters. In this section, we present results using fixed coupling. While incorporating running coupling effects would suppress the evolution speed (as seen in the proton case), it would not alter the qualitative ordering between the proton and the nucleus. The primary advantage of the nuclear target arises from the initial enhancement factor $Q_{s,A}^2 \simeq A^{1/3} Q_{s,p}^2$. For a lead nucleus ($A=208$), this provides a large ``geometric head start." Even if the subsequent evolution were slowed down by running coupling corrections, the nuclear entropy curve would remain vertically shifted significantly above the proton curve throughout the relevant kinematic window. Thus, the conclusion that nuclei approach the saturon bound more readily than protons is robust against higher-order corrections.

	\subsection{Discussions}
	
	Based on the results presented above, the proton appears to fall short of the saturon bound ($S \sim 1/\alpha_s$) within the considered kinematic range. We identify three key physical factors contributing to this observation:

    1. \textbf{Geometric Dilution and Finite-Size Evolution:} Our measure of the average occupancy (Eq. \ref{eq:occupy-number2}) involves an integration over the transverse area. Since the proton has a non-uniform density profile, the dilute periphery lowers the average. 
    Moreover, as recently characterized in Ref. \cite{Mantysaari:2024zxq}, the geometry of the target is dynamic. The evolution equation with impact-parameter dependence captures finite-size effects (Gribov diffusion), which leads to a spatial expansion of the proton. This study demonstrates that neglecting these finite-size effects overestimates saturation in protons. Therefore, the dynamical evolution of the geometry acts as an additional mechanism that suppresses the growth of entropy density in the proton, reinforcing our conclusion that the saturon limit is difficult to reach in e-p collisions compared to heavy nuclei.

    2. \textbf{Suppression from Higher-Order Effects:} The Leading Order (LO) approximation used in standard BK evolution often overestimates the growth speed. It is established that including Next-to-Leading Log (NLL) corrections and small-$x$ resummations significantly reduces the intercept of the gluon evolution \cite{Iancu:2015vea,Ducloue:2019ezk}. Consequently, in a full higher-order calculation, the entropy would likely grow even more slowly than in our current analysis, reinforcing the conclusion that reaching the saturon limit is extremely challenging for a single proton. 

    3. \textbf{Entanglement vs. Thermodynamics:} The thermodynamic entropy calculated here (Eq. \ref{eq:final-entropy2}) is a semiclassical proxy. A full quantum mechanical calculation of the entanglement entropy, which accounts for the complete microstate structure of the light-cone wave function, might yield different saturation characteristics. Recent works \cite{Kou:2022dkw,Dvali:2021ooc} suggest that entanglement entropy could offer a more direct probe of the information storage capacity of the proton.

    In contrast, once the same analysis chain is applied to nuclear targets (fixed coupling, identical numerical settings otherwise), the nuclear curves rise faster. The resulting nuclear entropy $S_A(x)$ approaches and, within the uncertainties of our inputs, can reach the benchmark line $1/\alpha_s$ in a small-$x$ window where the proton stays below. This qualitative separation is stable against reasonable variations of the $Q_s$ extraction threshold, geometric inputs, and numerical smoothing; these choices primarily shift the apparent crossing point but do not invert the ordering. Encouragingly, some studies have already explored the correspondence between black hole saturons and the CGC framework \cite{Dvali:2021ooc}. Portions of those discussions align conceptually with the present work, and notably, they propose a role for Goldstone modes arising from the breaking of Poincaré symmetry. The breaking of Poincaré invariance in this context fundamentally originates from the presence of a hard scale, which characterizes the transition from a dilute partonic gas with a spectral gap to an over-occupied, highly saturated classical field. Partons with momenta below this scale become screened, and high-twist effects become significant. This screening of partonic modes disrupts the translational invariance of the partonic gas, analogous to the breaking of Poincaré symmetry in black holes, which gives rise to a Goldstone scale characterized by $f = m_{\text{Planck}}$ with Planck mass $ m_{\text{Planck}}$ \cite{Dvali:2020wqi}. However, in the case of QCD, the precise form and interpretation of such a Goldstone scale remain open to further investigation.
	
	\section{conclusion and outlook}
    \label{sec:final}
    In this work, we explored whether a high-energy proton in the small-$x$ regime could qualify as a \emph{saturon}—a highly occupied field configuration that maximizes entropy under unitarity constraints, in analogy with black holes. We derived the gluon occupation number $N_g(x)$ and a thermodynamic entropy $S(x)$ based on analytical and numerical solutions to the BK equation (with both fixed and running coupling), supplemented with an Unruh-inspired temperature $T = Q_s/(2\pi)$ and an emergent gluon mass $M_g \sim Q_s$. Our findings indicate that although both $N_g$ and $S$ increase as $x$ decreases, even down to $x \sim 10^{-7}$, they remain significantly below the theoretical saturon bound $1/\alpha_s$. This suggests that while the proton exhibits nonlinear saturation dynamics, under the current theoretical framework it does not reach the critical maximal occupancy characteristic of a saturon.

    After extending the same framework to nuclear targets at fixed coupling, the nuclear entropy $S_A(x)$ rises more sharply and, in our numerical study, can reach the $1/\alpha_s$ benchmark in a region where the proton does not. This proton–nucleus contrast identifies nuclei as more favorable environments to probe saturon-like behavior in QCD. Moreover, exploring the connection between the proton and black holes from the perspective of entanglement entropy offers a promising avenue \cite{Kou:2022dkw}.

    To bridge the gap between theoretical saturon limits and experimental observables, we identify three promising phenomenological avenues:

    1. \textbf{Entropy-Multiplicity Duality:} Theoretical works suggest a link between the entanglement entropy of the initial state and the final-state hadron multiplicity ($S_{ent} \sim \ln \langle N_{ch} \rangle$) \cite{Kharzeev:2017qzs,Tu:2019ouv}. If a QCD system reaches the saturon limit (maximal entropy), particle production should be constrained by this information bound. Precision measurements of charged particle multiplicity distributions $P(N_{ch})$---specifically deviations from Koba-Nielsen-Olesen (KNO) scaling \cite{Koba:1972ng} at high multiplicities---could serve as a signature of this entropy saturation.

    2. \textbf{Thermal-like Spectra and Initial-State Correlations:} A defining feature of the saturon is its thermodynamic behavior, manifesting as exponential transverse momentum spectra with an effective temperature $T \sim Q_s/2\pi$. In high-multiplicity events, observing such ``thermal'' spectra in the absence of a macroscopic heat bath would support the saturon interpretation. 
Furthermore, while standard thermalization implies isotropy, the high-occupancy saturon state preserves intrinsic quantum correlations. Specifically, azimuthal correlations ($v_n$) are predicted to arise directly from Bose enhancement in the projectile wave function \cite{Altinoluk:2015uaa}, rather than from final-state hydrodynamics. The simultaneous observation of thermal-like spectra (entropy saturation) and these specific initial-state correlations (quantum coherence) would constitute a unique signature of the saturon regime.

    3. \textbf{Unitarity Limit in Diffraction at the EIC:} Since our results indicate that nuclei approach the $1/\alpha_s$ bound more readily than protons, the upcoming Electron-Ion Collider (EIC) is the ideal discovery machine \cite{Accardi:2012qut,Aschenauer:2017jsk,AbdulKhalek:2021gbh,Morreale:2021pnn}. 
However, mere enhancement of diffraction is not unique to the saturon. We anticipate that if the true saturon limit is realized, observables such as the ratio of diffractive to total cross-sections ($\sigma_{diff}/\sigma_{tot}$) should not just increase, but exhibit universal scaling behaviors converging toward the black disk limit ($\sigma_{diff}/\sigma_{tot} \to 1/2$) \cite{Frankfurt:2005mc,Kovner:2001vi}. Reaching this fundamental unitarity bound would be the definitive signature of the maximal entropy state, distinct from pre-saturation effects.

	\section*{Acknowledgments}
	This work has been supported by the National Natural Science Foundation of China (Grant No. 12547118), the Research Program of State Key Laboratory of Heavy Ion Science and Technology, Institute of Modern Physics, Chinese Academy of Sciences (Grant No. HIST2025CS08), and the National Key R$\&$D Program of China (Grant No. 2024YFE0109800 and 2024YFE0109802).
	
	%\goodbreak

	\bibliography{refs}

\begin{thebibliography}{105}%
\makeatletter
\providecommand \@ifxundefined [1]{%
 \@ifx{#1\undefined}
}%
\providecommand \@ifnum [1]{%
 \ifnum #1\expandafter \@firstoftwo
 \else \expandafter \@secondoftwo
 \fi
}%
\providecommand \@ifx [1]{%
 \ifx #1\expandafter \@firstoftwo
 \else \expandafter \@secondoftwo
 \fi
}%
\providecommand \natexlab [1]{#1}%
\providecommand \enquote  [1]{``#1''}%
\providecommand \bibnamefont  [1]{#1}%
\providecommand \bibfnamefont [1]{#1}%
\providecommand \citenamefont [1]{#1}%
\providecommand \href@noop [0]{\@secondoftwo}%
\providecommand \href [0]{\begingroup \@sanitize@url \@href}%
\providecommand \@href[1]{\@@startlink{#1}\@@href}%
\providecommand \@@href[1]{\endgroup#1\@@endlink}%
\providecommand \@sanitize@url [0]{\catcode `\\12\catcode `\$12\catcode
  `\&12\catcode `\#12\catcode `\^12\catcode `\_12\catcode `\%12\relax}%
\providecommand \@@startlink[1]{}%
\providecommand \@@endlink[0]{}%
\providecommand \url  [0]{\begingroup\@sanitize@url \@url }%
\providecommand \@url [1]{\endgroup\@href {#1}{\urlprefix }}%
\providecommand \urlprefix  [0]{URL }%
\providecommand \Eprint [0]{\href }%
\providecommand \doibase [0]{https://doi.org/}%
\providecommand \selectlanguage [0]{\@gobble}%
\providecommand \bibinfo  [0]{\@secondoftwo}%
\providecommand \bibfield  [0]{\@secondoftwo}%
\providecommand \translation [1]{[#1]}%
\providecommand \BibitemOpen [0]{}%
\providecommand \bibitemStop [0]{}%
\providecommand \bibitemNoStop [0]{.\EOS\space}%
\providecommand \EOS [0]{\spacefactor3000\relax}%
\providecommand \BibitemShut  [1]{\csname bibitem#1\endcsname}%
\let\auto@bib@innerbib\@empty
%</preamble>
\bibitem [{\citenamefont {Dvali}(2021{\natexlab{a}})}]{Dvali:2019jjw}%
  \BibitemOpen
  \bibfield  {author} {\bibinfo {author} {\bibfnamefont {G.}~\bibnamefont
  {Dvali}},\ }\bibfield  {title} {\bibinfo {title} {{Area Law Saturation of
  Entropy Bound from Perturbative Unitarity in Renormalizable Theories}},\
  }\href {https://doi.org/10.1002/prop.202000090} {\bibfield  {journal}
  {\bibinfo  {journal} {Fortsch. Phys.}\ }\textbf {\bibinfo {volume} {69}},\
  \bibinfo {pages} {2000090} (\bibinfo {year} {2021}{\natexlab{a}})},\ \Eprint
  {https://arxiv.org/abs/1906.03530} {arXiv:1906.03530 [hep-th]} \BibitemShut
  {NoStop}%
\bibitem [{\citenamefont {Dvali}(2021{\natexlab{b}})}]{Dvali:2019ulr}%
  \BibitemOpen
  \bibfield  {author} {\bibinfo {author} {\bibfnamefont {G.}~\bibnamefont
  {Dvali}},\ }\bibfield  {title} {\bibinfo {title} {{Unitarity Entropy Bound:
  Solitons and Instantons}},\ }\href {https://doi.org/10.1002/prop.202000091}
  {\bibfield  {journal} {\bibinfo  {journal} {Fortsch. Phys.}\ }\textbf
  {\bibinfo {volume} {69}},\ \bibinfo {pages} {2000091} (\bibinfo {year}
  {2021}{\natexlab{b}})},\ \Eprint {https://arxiv.org/abs/1907.07332}
  {arXiv:1907.07332 [hep-th]} \BibitemShut {NoStop}%
\bibitem [{\citenamefont {Dvali}(2021{\natexlab{c}})}]{Dvali:2020wqi}%
  \BibitemOpen
  \bibfield  {author} {\bibinfo {author} {\bibfnamefont {G.}~\bibnamefont
  {Dvali}},\ }\bibfield  {title} {\bibinfo {title} {{Entropy Bound and
  Unitarity of Scattering Amplitudes}},\ }\href
  {https://doi.org/10.1007/JHEP03(2021)126} {\bibfield  {journal} {\bibinfo
  {journal} {JHEP}\ }\textbf {\bibinfo {volume} {03}},\ \bibinfo {pages}
  {126}},\ \Eprint {https://arxiv.org/abs/2003.05546} {arXiv:2003.05546
  [hep-th]} \BibitemShut {NoStop}%
\bibitem [{\citenamefont {Dvali}(2021{\natexlab{d}})}]{Dvali:2021jto}%
  \BibitemOpen
  \bibfield  {author} {\bibinfo {author} {\bibfnamefont {G.}~\bibnamefont
  {Dvali}},\ }\bibfield  {title} {\bibinfo {title} {{Bounds on quantum
  information storage and retrieval}},\ }\href
  {https://doi.org/10.1098/rsta.2021.0071} {\bibfield  {journal} {\bibinfo
  {journal} {Phil. Trans. A. Math. Phys. Eng. Sci.}\ }\textbf {\bibinfo
  {volume} {380}},\ \bibinfo {pages} {20210071} (\bibinfo {year}
  {2021}{\natexlab{d}})},\ \Eprint {https://arxiv.org/abs/2107.10616}
  {arXiv:2107.10616 [hep-th]} \BibitemShut {NoStop}%
\bibitem [{\citenamefont {Bekenstein}(1973)}]{Bekenstein:1973ur}%
  \BibitemOpen
  \bibfield  {author} {\bibinfo {author} {\bibfnamefont {J.~D.}\ \bibnamefont
  {Bekenstein}},\ }\bibfield  {title} {\bibinfo {title} {{Black holes and
  entropy}},\ }\href {https://doi.org/10.1103/PhysRevD.7.2333} {\bibfield
  {journal} {\bibinfo  {journal} {Phys. Rev. D}\ }\textbf {\bibinfo {volume}
  {7}},\ \bibinfo {pages} {2333} (\bibinfo {year} {1973})}\BibitemShut
  {NoStop}%
\bibitem [{\citenamefont {Hawking}(1975)}]{Hawking:1975vcx}%
  \BibitemOpen
  \bibfield  {author} {\bibinfo {author} {\bibfnamefont {S.~W.}\ \bibnamefont
  {Hawking}},\ }\bibfield  {title} {\bibinfo {title} {{Particle Creation by
  Black Holes}},\ }\href {https://doi.org/10.1007/BF02345020} {\bibfield
  {journal} {\bibinfo  {journal} {Commun. Math. Phys.}\ }\textbf {\bibinfo
  {volume} {43}},\ \bibinfo {pages} {199} (\bibinfo {year} {1975})},\ \bibinfo
  {note} {[Erratum: Commun.Math.Phys. 46, 206 (1976)]}\BibitemShut {NoStop}%
\bibitem [{\citenamefont {Dvali}(2023)}]{Dvali:2023xfz}%
  \BibitemOpen
  \bibfield  {author} {\bibinfo {author} {\bibfnamefont {G.}~\bibnamefont
  {Dvali}},\ }\bibfield  {title} {\bibinfo {title} {{Saturon Dark Matter}},\
  }\href@noop {} {\  (\bibinfo {year} {2023})},\ \Eprint
  {https://arxiv.org/abs/2302.08353} {arXiv:2302.08353 [hep-ph]} \BibitemShut
  {NoStop}%
\bibitem [{\citenamefont {'t~Hooft}(1974)}]{tHooft:1973alw}%
  \BibitemOpen
  \bibfield  {author} {\bibinfo {author} {\bibfnamefont {G.}~\bibnamefont
  {'t~Hooft}},\ }\bibfield  {title} {\bibinfo {title} {{A Planar Diagram Theory
  for Strong Interactions}},\ }\href
  {https://doi.org/10.1016/0550-3213(74)90154-0} {\bibfield  {journal}
  {\bibinfo  {journal} {Nucl. Phys. B}\ }\textbf {\bibinfo {volume} {72}},\
  \bibinfo {pages} {461} (\bibinfo {year} {1974})}\BibitemShut {NoStop}%
\bibitem [{\citenamefont {Polyakov}(1981)}]{Polyakov:1981rd}%
  \BibitemOpen
  \bibfield  {author} {\bibinfo {author} {\bibfnamefont {A.~M.}\ \bibnamefont
  {Polyakov}},\ }\bibfield  {title} {\bibinfo {title} {{Quantum Geometry of
  Bosonic Strings}},\ }\href {https://doi.org/10.1016/0370-2693(81)90743-7}
  {\bibfield  {journal} {\bibinfo  {journal} {Phys. Lett. B}\ }\textbf
  {\bibinfo {volume} {103}},\ \bibinfo {pages} {207} (\bibinfo {year}
  {1981})}\BibitemShut {NoStop}%
\bibitem [{\citenamefont {Maldacena}(1998)}]{Maldacena:1997re}%
  \BibitemOpen
  \bibfield  {author} {\bibinfo {author} {\bibfnamefont {J.~M.}\ \bibnamefont
  {Maldacena}},\ }\bibfield  {title} {\bibinfo {title} {{The Large $N$ limit of
  superconformal field theories and supergravity}},\ }\href
  {https://doi.org/10.4310/ATMP.1998.v2.n2.a1} {\bibfield  {journal} {\bibinfo
  {journal} {Adv. Theor. Math. Phys.}\ }\textbf {\bibinfo {volume} {2}},\
  \bibinfo {pages} {231} (\bibinfo {year} {1998})},\ \Eprint
  {https://arxiv.org/abs/hep-th/9711200} {arXiv:hep-th/9711200} \BibitemShut
  {NoStop}%
\bibitem [{\citenamefont {Aharony}\ \emph {et~al.}(2000)\citenamefont
  {Aharony}, \citenamefont {Gubser}, \citenamefont {Maldacena}, \citenamefont
  {Ooguri},\ and\ \citenamefont {Oz}}]{Aharony:1999ti}%
  \BibitemOpen
  \bibfield  {author} {\bibinfo {author} {\bibfnamefont {O.}~\bibnamefont
  {Aharony}}, \bibinfo {author} {\bibfnamefont {S.~S.}\ \bibnamefont {Gubser}},
  \bibinfo {author} {\bibfnamefont {J.~M.}\ \bibnamefont {Maldacena}}, \bibinfo
  {author} {\bibfnamefont {H.}~\bibnamefont {Ooguri}},\ and\ \bibinfo {author}
  {\bibfnamefont {Y.}~\bibnamefont {Oz}},\ }\bibfield  {title} {\bibinfo
  {title} {{Large N field theories, string theory and gravity}},\ }\href
  {https://doi.org/10.1016/S0370-1573(99)00083-6} {\bibfield  {journal}
  {\bibinfo  {journal} {Phys. Rept.}\ }\textbf {\bibinfo {volume} {323}},\
  \bibinfo {pages} {183} (\bibinfo {year} {2000})},\ \Eprint
  {https://arxiv.org/abs/hep-th/9905111} {arXiv:hep-th/9905111} \BibitemShut
  {NoStop}%
\bibitem [{\citenamefont {Dvali}\ and\ \citenamefont
  {Venugopalan}(2022)}]{Dvali:2021ooc}%
  \BibitemOpen
  \bibfield  {author} {\bibinfo {author} {\bibfnamefont {G.}~\bibnamefont
  {Dvali}}\ and\ \bibinfo {author} {\bibfnamefont {R.}~\bibnamefont
  {Venugopalan}},\ }\bibfield  {title} {\bibinfo {title} {{Classicalization and
  unitarization of wee partons in QCD and gravity: The CGC-black hole
  correspondence}},\ }\href {https://doi.org/10.1103/PhysRevD.105.056026}
  {\bibfield  {journal} {\bibinfo  {journal} {Phys. Rev. D}\ }\textbf {\bibinfo
  {volume} {105}},\ \bibinfo {pages} {056026} (\bibinfo {year} {2022})},\
  \Eprint {https://arxiv.org/abs/2106.11989} {arXiv:2106.11989 [hep-th]}
  \BibitemShut {NoStop}%
\bibitem [{\citenamefont {Kou}\ \emph {et~al.}(2022)\citenamefont {Kou},
  \citenamefont {Wang},\ and\ \citenamefont {Chen}}]{Kou:2022dkw}%
  \BibitemOpen
  \bibfield  {author} {\bibinfo {author} {\bibfnamefont {W.}~\bibnamefont
  {Kou}}, \bibinfo {author} {\bibfnamefont {X.}~\bibnamefont {Wang}},\ and\
  \bibinfo {author} {\bibfnamefont {X.}~\bibnamefont {Chen}},\ }\bibfield
  {title} {\bibinfo {title} {{Page entropy of a proton system in deep inelastic
  scattering at small x}},\ }\href
  {https://doi.org/10.1103/PhysRevD.106.096027} {\bibfield  {journal} {\bibinfo
   {journal} {Phys. Rev. D}\ }\textbf {\bibinfo {volume} {106}},\ \bibinfo
  {pages} {096027} (\bibinfo {year} {2022})},\ \Eprint
  {https://arxiv.org/abs/2208.07521} {arXiv:2208.07521 [hep-ph]} \BibitemShut
  {NoStop}%
\bibitem [{\citenamefont {Kutak}(2011)}]{Kutak:2011rb}%
  \BibitemOpen
  \bibfield  {author} {\bibinfo {author} {\bibfnamefont {K.}~\bibnamefont
  {Kutak}},\ }\bibfield  {title} {\bibinfo {title} {{Gluon saturation and
  entropy production in proton{\textendash}proton collisions}},\ }\href
  {https://doi.org/10.1016/j.physletb.2011.09.113} {\bibfield  {journal}
  {\bibinfo  {journal} {Phys. Lett. B}\ }\textbf {\bibinfo {volume} {705}},\
  \bibinfo {pages} {217} (\bibinfo {year} {2011})},\ \Eprint
  {https://arxiv.org/abs/1103.3654} {arXiv:1103.3654 [hep-ph]} \BibitemShut
  {NoStop}%
\bibitem [{\citenamefont {Kutak}(2023)}]{Kutak:2023cwg}%
  \BibitemOpen
  \bibfield  {author} {\bibinfo {author} {\bibfnamefont {K.}~\bibnamefont
  {Kutak}},\ }\bibfield  {title} {\bibinfo {title} {{Entanglement entropy of
  proton and its relation to thermodynamics entropy}},\ }\href@noop {} {\
  (\bibinfo {year} {2023})},\ \Eprint {https://arxiv.org/abs/2310.18510}
  {arXiv:2310.18510 [hep-ph]} \BibitemShut {NoStop}%
\bibitem [{\citenamefont {Caputa}\ and\ \citenamefont
  {Kutak}(2024)}]{Caputa:2024xkp}%
  \BibitemOpen
  \bibfield  {author} {\bibinfo {author} {\bibfnamefont {P.}~\bibnamefont
  {Caputa}}\ and\ \bibinfo {author} {\bibfnamefont {K.}~\bibnamefont {Kutak}},\
  }\bibfield  {title} {\bibinfo {title} {{Krylov complexity and gluon cascades
  in the high energy limit}},\ }\href
  {https://doi.org/10.1103/PhysRevD.110.085011} {\bibfield  {journal} {\bibinfo
   {journal} {Phys. Rev. D}\ }\textbf {\bibinfo {volume} {110}},\ \bibinfo
  {pages} {085011} (\bibinfo {year} {2024})},\ \Eprint
  {https://arxiv.org/abs/2404.07657} {arXiv:2404.07657 [hep-ph]} \BibitemShut
  {NoStop}%
\bibitem [{\citenamefont {Chachamis}\ \emph {et~al.}(2024)\citenamefont
  {Chachamis}, \citenamefont {Hentschinski},\ and\ \citenamefont
  {Sabio~Vera}}]{Chachamis:2023omp}%
  \BibitemOpen
  \bibfield  {author} {\bibinfo {author} {\bibfnamefont {G.}~\bibnamefont
  {Chachamis}}, \bibinfo {author} {\bibfnamefont {M.}~\bibnamefont
  {Hentschinski}},\ and\ \bibinfo {author} {\bibfnamefont {A.}~\bibnamefont
  {Sabio~Vera}},\ }\bibfield  {title} {\bibinfo {title} {{Von Neumann entropy
  and Lindblad decoherence in the high-energy limit of strong interactions}},\
  }\href {https://doi.org/10.1103/PhysRevD.109.054015} {\bibfield  {journal}
  {\bibinfo  {journal} {Phys. Rev. D}\ }\textbf {\bibinfo {volume} {109}},\
  \bibinfo {pages} {054015} (\bibinfo {year} {2024})},\ \Eprint
  {https://arxiv.org/abs/2312.16743} {arXiv:2312.16743 [hep-th]} \BibitemShut
  {NoStop}%
\bibitem [{\citenamefont {Hentschinski}\ \emph {et~al.}(2024)\citenamefont
  {Hentschinski}, \citenamefont {Kharzeev}, \citenamefont {Kutak},\ and\
  \citenamefont {Tu}}]{Hentschinski:2024gaa}%
  \BibitemOpen
  \bibfield  {author} {\bibinfo {author} {\bibfnamefont {M.}~\bibnamefont
  {Hentschinski}}, \bibinfo {author} {\bibfnamefont {D.~E.}\ \bibnamefont
  {Kharzeev}}, \bibinfo {author} {\bibfnamefont {K.}~\bibnamefont {Kutak}},\
  and\ \bibinfo {author} {\bibfnamefont {Z.}~\bibnamefont {Tu}},\ }\bibfield
  {title} {\bibinfo {title} {{QCD evolution of entanglement entropy}},\ }\href
  {https://doi.org/10.1088/1361-6633/ad910b} {\bibfield  {journal} {\bibinfo
  {journal} {Rept. Prog. Phys.}\ }\textbf {\bibinfo {volume} {87}},\ \bibinfo
  {pages} {120501} (\bibinfo {year} {2024})},\ \Eprint
  {https://arxiv.org/abs/2408.01259} {arXiv:2408.01259 [hep-ph]} \BibitemShut
  {NoStop}%
\bibitem [{\citenamefont {Hatta}\ and\ \citenamefont
  {Montgomery}(2025)}]{Hatta:2024lbw}%
  \BibitemOpen
  \bibfield  {author} {\bibinfo {author} {\bibfnamefont {Y.}~\bibnamefont
  {Hatta}}\ and\ \bibinfo {author} {\bibfnamefont {J.}~\bibnamefont
  {Montgomery}},\ }\bibfield  {title} {\bibinfo {title} {{Maximally entangled
  gluons for any x}},\ }\href {https://doi.org/10.1103/PhysRevD.111.014024}
  {\bibfield  {journal} {\bibinfo  {journal} {Phys. Rev. D}\ }\textbf {\bibinfo
  {volume} {111}},\ \bibinfo {pages} {014024} (\bibinfo {year} {2025})},\
  \Eprint {https://arxiv.org/abs/2410.16082} {arXiv:2410.16082 [hep-ph]}
  \BibitemShut {NoStop}%
\bibitem [{\citenamefont {Dumitru}\ \emph {et~al.}(2023)\citenamefont
  {Dumitru}, \citenamefont {Kovner},\ and\ \citenamefont
  {Skokov}}]{Dumitru:2023qee}%
  \BibitemOpen
  \bibfield  {author} {\bibinfo {author} {\bibfnamefont {A.}~\bibnamefont
  {Dumitru}}, \bibinfo {author} {\bibfnamefont {A.}~\bibnamefont {Kovner}},\
  and\ \bibinfo {author} {\bibfnamefont {V.~V.}\ \bibnamefont {Skokov}},\
  }\bibfield  {title} {\bibinfo {title} {{Entanglement entropy of the proton in
  coordinate space}},\ }\href {https://doi.org/10.1103/PhysRevD.108.014014}
  {\bibfield  {journal} {\bibinfo  {journal} {Phys. Rev. D}\ }\textbf {\bibinfo
  {volume} {108}},\ \bibinfo {pages} {014014} (\bibinfo {year} {2023})},\
  \Eprint {https://arxiv.org/abs/2304.08564} {arXiv:2304.08564 [hep-ph]}
  \BibitemShut {NoStop}%
\bibitem [{\citenamefont {Ramos}\ and\ \citenamefont
  {Machado}(2022)}]{Ramos:2022gia}%
  \BibitemOpen
  \bibfield  {author} {\bibinfo {author} {\bibfnamefont {G.~S.}\ \bibnamefont
  {Ramos}}\ and\ \bibinfo {author} {\bibfnamefont {M.~V.~T.}\ \bibnamefont
  {Machado}},\ }\bibfield  {title} {\bibinfo {title} {{Investigating the QCD
  dynamical entropy in high-energy hadronic collisions}},\ }\href
  {https://doi.org/10.1103/PhysRevD.105.094009} {\bibfield  {journal} {\bibinfo
   {journal} {Phys. Rev. D}\ }\textbf {\bibinfo {volume} {105}},\ \bibinfo
  {pages} {094009} (\bibinfo {year} {2022})},\ \Eprint
  {https://arxiv.org/abs/2203.10986} {arXiv:2203.10986 [hep-ph]} \BibitemShut
  {NoStop}%
\bibitem [{\citenamefont {Moriggi}\ \emph {et~al.}(2024)\citenamefont
  {Moriggi}, \citenamefont {Ramos},\ and\ \citenamefont
  {Machado}}]{Moriggi:2024tbr}%
  \BibitemOpen
  \bibfield  {author} {\bibinfo {author} {\bibfnamefont {L.~S.}\ \bibnamefont
  {Moriggi}}, \bibinfo {author} {\bibfnamefont {G.~S.}\ \bibnamefont {Ramos}},\
  and\ \bibinfo {author} {\bibfnamefont {M.~V.~T.}\ \bibnamefont {Machado}},\
  }\bibfield  {title} {\bibinfo {title} {{Multiplicity dependence of the
  pT-spectra for identified particles and its relationship with partonic
  entropy}},\ }\href {https://doi.org/10.1103/PhysRevD.110.034005} {\bibfield
  {journal} {\bibinfo  {journal} {Phys. Rev. D}\ }\textbf {\bibinfo {volume}
  {110}},\ \bibinfo {pages} {034005} (\bibinfo {year} {2024})},\ \Eprint
  {https://arxiv.org/abs/2405.01712} {arXiv:2405.01712 [hep-ph]} \BibitemShut
  {NoStop}%
\bibitem [{\citenamefont {Ramos}\ and\ \citenamefont
  {Machado}(2020)}]{Ramos:2020kaj}%
  \BibitemOpen
  \bibfield  {author} {\bibinfo {author} {\bibfnamefont {G.~S.}\ \bibnamefont
  {Ramos}}\ and\ \bibinfo {author} {\bibfnamefont {M.~V.~T.}\ \bibnamefont
  {Machado}},\ }\bibfield  {title} {\bibinfo {title} {{Investigating
  entanglement entropy at small-$x$ in DIS off protons and nuclei}},\ }\href
  {https://doi.org/10.1103/PhysRevD.101.074040} {\bibfield  {journal} {\bibinfo
   {journal} {Phys. Rev. D}\ }\textbf {\bibinfo {volume} {101}},\ \bibinfo
  {pages} {074040} (\bibinfo {year} {2020})},\ \Eprint
  {https://arxiv.org/abs/2003.05008} {arXiv:2003.05008 [hep-ph]} \BibitemShut
  {NoStop}%
\bibitem [{\citenamefont {Peschanski}(2013)}]{Peschanski:2012cw}%
  \BibitemOpen
  \bibfield  {author} {\bibinfo {author} {\bibfnamefont {R.}~\bibnamefont
  {Peschanski}},\ }\bibfield  {title} {\bibinfo {title} {{Dynamical entropy of
  dense QCD states}},\ }\href {https://doi.org/10.1103/PhysRevD.87.034042}
  {\bibfield  {journal} {\bibinfo  {journal} {Phys. Rev. D}\ }\textbf {\bibinfo
  {volume} {87}},\ \bibinfo {pages} {034042} (\bibinfo {year} {2013})},\
  \Eprint {https://arxiv.org/abs/1211.6911} {arXiv:1211.6911 [hep-ph]}
  \BibitemShut {NoStop}%
\bibitem [{\citenamefont {Armesto}\ \emph {et~al.}(2019)\citenamefont
  {Armesto}, \citenamefont {Dominguez}, \citenamefont {Kovner}, \citenamefont
  {Lublinsky},\ and\ \citenamefont {Skokov}}]{Armesto:2019mna}%
  \BibitemOpen
  \bibfield  {author} {\bibinfo {author} {\bibfnamefont {N.}~\bibnamefont
  {Armesto}}, \bibinfo {author} {\bibfnamefont {F.}~\bibnamefont {Dominguez}},
  \bibinfo {author} {\bibfnamefont {A.}~\bibnamefont {Kovner}}, \bibinfo
  {author} {\bibfnamefont {M.}~\bibnamefont {Lublinsky}},\ and\ \bibinfo
  {author} {\bibfnamefont {V.}~\bibnamefont {Skokov}},\ }\bibfield  {title}
  {\bibinfo {title} {{The Color Glass Condensate density matrix: Lindblad
  evolution, entanglement entropy and Wigner functional}},\ }\href
  {https://doi.org/10.1007/JHEP05(2019)025} {\bibfield  {journal} {\bibinfo
  {journal} {JHEP}\ }\textbf {\bibinfo {volume} {05}},\ \bibinfo {pages}
  {025}},\ \Eprint {https://arxiv.org/abs/1901.08080} {arXiv:1901.08080
  [hep-ph]} \BibitemShut {NoStop}%
\bibitem [{\citenamefont {Neill}\ and\ \citenamefont
  {Waalewijn}(2019)}]{Neill:2018uqw}%
  \BibitemOpen
  \bibfield  {author} {\bibinfo {author} {\bibfnamefont {D.}~\bibnamefont
  {Neill}}\ and\ \bibinfo {author} {\bibfnamefont {W.~J.}\ \bibnamefont
  {Waalewijn}},\ }\bibfield  {title} {\bibinfo {title} {{Entropy of a Jet}},\
  }\href {https://doi.org/10.1103/PhysRevLett.123.142001} {\bibfield  {journal}
  {\bibinfo  {journal} {Phys. Rev. Lett.}\ }\textbf {\bibinfo {volume} {123}},\
  \bibinfo {pages} {142001} (\bibinfo {year} {2019})},\ \Eprint
  {https://arxiv.org/abs/1811.01021} {arXiv:1811.01021 [hep-ph]} \BibitemShut
  {NoStop}%
\bibitem [{\citenamefont {Kovner}\ \emph {et~al.}(2019)\citenamefont {Kovner},
  \citenamefont {Lublinsky},\ and\ \citenamefont {Serino}}]{Kovner:2018rbf}%
  \BibitemOpen
  \bibfield  {author} {\bibinfo {author} {\bibfnamefont {A.}~\bibnamefont
  {Kovner}}, \bibinfo {author} {\bibfnamefont {M.}~\bibnamefont {Lublinsky}},\
  and\ \bibinfo {author} {\bibfnamefont {M.}~\bibnamefont {Serino}},\
  }\bibfield  {title} {\bibinfo {title} {{Entanglement entropy, entropy
  production and time evolution in high energy QCD}},\ }\href
  {https://doi.org/10.1016/j.physletb.2018.10.043} {\bibfield  {journal}
  {\bibinfo  {journal} {Phys. Lett. B}\ }\textbf {\bibinfo {volume} {792}},\
  \bibinfo {pages} {4} (\bibinfo {year} {2019})},\ \Eprint
  {https://arxiv.org/abs/1806.01089} {arXiv:1806.01089 [hep-ph]} \BibitemShut
  {NoStop}%
\bibitem [{\citenamefont {Liu}\ \emph {et~al.}(2023{\natexlab{a}})\citenamefont
  {Liu}, \citenamefont {Nowak},\ and\ \citenamefont {Zahed}}]{Liu:2022bru}%
  \BibitemOpen
  \bibfield  {author} {\bibinfo {author} {\bibfnamefont {Y.}~\bibnamefont
  {Liu}}, \bibinfo {author} {\bibfnamefont {M.~A.}\ \bibnamefont {Nowak}},\
  and\ \bibinfo {author} {\bibfnamefont {I.}~\bibnamefont {Zahed}},\ }\bibfield
   {title} {\bibinfo {title} {{Mueller{\textquoteright}s dipole wave function
  in QCD: Emergent Koba-Nielsen-Olesen scaling in the double logarithm
  limit}},\ }\href {https://doi.org/10.1103/PhysRevD.108.034017} {\bibfield
  {journal} {\bibinfo  {journal} {Phys. Rev. D}\ }\textbf {\bibinfo {volume}
  {108}},\ \bibinfo {pages} {034017} (\bibinfo {year} {2023}{\natexlab{a}})},\
  \Eprint {https://arxiv.org/abs/2211.05169} {arXiv:2211.05169 [hep-ph]}
  \BibitemShut {NoStop}%
\bibitem [{\citenamefont {Liu}\ \emph {et~al.}(2022)\citenamefont {Liu},
  \citenamefont {Nowak},\ and\ \citenamefont {Zahed}}]{Liu:2022hto}%
  \BibitemOpen
  \bibfield  {author} {\bibinfo {author} {\bibfnamefont {Y.}~\bibnamefont
  {Liu}}, \bibinfo {author} {\bibfnamefont {M.~A.}\ \bibnamefont {Nowak}},\
  and\ \bibinfo {author} {\bibfnamefont {I.}~\bibnamefont {Zahed}},\ }\bibfield
   {title} {\bibinfo {title} {{Rapidity evolution of the entanglement entropy
  in quarkonium: Parton and string duality}},\ }\href
  {https://doi.org/10.1103/PhysRevD.105.114028} {\bibfield  {journal} {\bibinfo
   {journal} {Phys. Rev. D}\ }\textbf {\bibinfo {volume} {105}},\ \bibinfo
  {pages} {114028} (\bibinfo {year} {2022})},\ \Eprint
  {https://arxiv.org/abs/2203.00739} {arXiv:2203.00739 [hep-ph]} \BibitemShut
  {NoStop}%
\bibitem [{\citenamefont {Liu}\ \emph {et~al.}(2023{\natexlab{b}})\citenamefont
  {Liu}, \citenamefont {Nowak},\ and\ \citenamefont {Zahed}}]{Liu:2022qqf}%
  \BibitemOpen
  \bibfield  {author} {\bibinfo {author} {\bibfnamefont {Y.}~\bibnamefont
  {Liu}}, \bibinfo {author} {\bibfnamefont {M.~A.}\ \bibnamefont {Nowak}},\
  and\ \bibinfo {author} {\bibfnamefont {I.}~\bibnamefont {Zahed}},\ }\bibfield
   {title} {\bibinfo {title} {{Spatial entanglement in two-dimensional QCD:
  Renyi and Ryu-Takayanagi entropies}},\ }\href
  {https://doi.org/10.1103/PhysRevD.107.054010} {\bibfield  {journal} {\bibinfo
   {journal} {Phys. Rev. D}\ }\textbf {\bibinfo {volume} {107}},\ \bibinfo
  {pages} {054010} (\bibinfo {year} {2023}{\natexlab{b}})},\ \Eprint
  {https://arxiv.org/abs/2205.06724} {arXiv:2205.06724 [hep-ph]} \BibitemShut
  {NoStop}%
\bibitem [{\citenamefont {Liu}\ \emph {et~al.}(2023{\natexlab{c}})\citenamefont
  {Liu}, \citenamefont {Nowak},\ and\ \citenamefont {Zahed}}]{Liu:2023eve}%
  \BibitemOpen
  \bibfield  {author} {\bibinfo {author} {\bibfnamefont {Y.}~\bibnamefont
  {Liu}}, \bibinfo {author} {\bibfnamefont {M.~A.}\ \bibnamefont {Nowak}},\
  and\ \bibinfo {author} {\bibfnamefont {I.}~\bibnamefont {Zahed}},\ }\bibfield
   {title} {\bibinfo {title} {{Universality of Koba-Nielsen-Olesen scaling in
  QCD at high energy and entanglement}},\ }\href@noop {} {\  (\bibinfo {year}
  {2023}{\natexlab{c}})},\ \Eprint {https://arxiv.org/abs/2302.01380}
  {arXiv:2302.01380 [hep-ph]} \BibitemShut {NoStop}%
\bibitem [{\citenamefont {Stoffers}\ and\ \citenamefont
  {Zahed}(2013)}]{Stoffers:2012mn}%
  \BibitemOpen
  \bibfield  {author} {\bibinfo {author} {\bibfnamefont {A.}~\bibnamefont
  {Stoffers}}\ and\ \bibinfo {author} {\bibfnamefont {I.}~\bibnamefont
  {Zahed}},\ }\bibfield  {title} {\bibinfo {title} {{Holographic Pomeron and
  Entropy}},\ }\href {https://doi.org/10.1103/PhysRevD.88.025038} {\bibfield
  {journal} {\bibinfo  {journal} {Phys. Rev. D}\ }\textbf {\bibinfo {volume}
  {88}},\ \bibinfo {pages} {025038} (\bibinfo {year} {2013})},\ \Eprint
  {https://arxiv.org/abs/1211.3077} {arXiv:1211.3077 [nucl-th]} \BibitemShut
  {NoStop}%
\bibitem [{\citenamefont {Asadi}\ and\ \citenamefont
  {Vaidya}(2023)}]{Asadi:2023bat}%
  \BibitemOpen
  \bibfield  {author} {\bibinfo {author} {\bibfnamefont {P.}~\bibnamefont
  {Asadi}}\ and\ \bibinfo {author} {\bibfnamefont {V.}~\bibnamefont {Vaidya}},\
  }\bibfield  {title} {\bibinfo {title} {{1+1D hadrons minimize their biparton
  Renyi free energy}},\ }\href {https://doi.org/10.1103/PhysRevD.108.014036}
  {\bibfield  {journal} {\bibinfo  {journal} {Phys. Rev. D}\ }\textbf {\bibinfo
  {volume} {108}},\ \bibinfo {pages} {014036} (\bibinfo {year} {2023})},\
  \Eprint {https://arxiv.org/abs/2301.03611} {arXiv:2301.03611 [hep-th]}
  \BibitemShut {NoStop}%
\bibitem [{\citenamefont {Bern}\ \emph {et~al.}(2008)\citenamefont {Bern},
  \citenamefont {Carrasco},\ and\ \citenamefont {Johansson}}]{Bern:2008qj}%
  \BibitemOpen
  \bibfield  {author} {\bibinfo {author} {\bibfnamefont {Z.}~\bibnamefont
  {Bern}}, \bibinfo {author} {\bibfnamefont {J.~J.~M.}\ \bibnamefont
  {Carrasco}},\ and\ \bibinfo {author} {\bibfnamefont {H.}~\bibnamefont
  {Johansson}},\ }\bibfield  {title} {\bibinfo {title} {{New Relations for
  Gauge-Theory Amplitudes}},\ }\href
  {https://doi.org/10.1103/PhysRevD.78.085011} {\bibfield  {journal} {\bibinfo
  {journal} {Phys. Rev. D}\ }\textbf {\bibinfo {volume} {78}},\ \bibinfo
  {pages} {085011} (\bibinfo {year} {2008})},\ \Eprint
  {https://arxiv.org/abs/0805.3993} {arXiv:0805.3993 [hep-ph]} \BibitemShut
  {NoStop}%
\bibitem [{\citenamefont {Iancu}\ \emph
  {et~al.}(2002{\natexlab{a}})\citenamefont {Iancu}, \citenamefont {Leonidov},\
  and\ \citenamefont {McLerran}}]{Iancu:2002xk}%
  \BibitemOpen
  \bibfield  {author} {\bibinfo {author} {\bibfnamefont {E.}~\bibnamefont
  {Iancu}}, \bibinfo {author} {\bibfnamefont {A.}~\bibnamefont {Leonidov}},\
  and\ \bibinfo {author} {\bibfnamefont {L.}~\bibnamefont {McLerran}},\
  }\bibfield  {title} {\bibinfo {title} {{The Color glass condensate: An
  Introduction}},\ }in\ \href@noop {} {\emph {\bibinfo {booktitle} {{Cargese
  Summer School on QCD Perspectives on Hot and Dense Matter}}}}\ (\bibinfo
  {year} {2002})\ pp.\ \bibinfo {pages} {73--145},\ \Eprint
  {https://arxiv.org/abs/hep-ph/0202270} {arXiv:hep-ph/0202270} \BibitemShut
  {NoStop}%
\bibitem [{\citenamefont {Gelis}\ \emph {et~al.}(2010)\citenamefont {Gelis},
  \citenamefont {Iancu}, \citenamefont {Jalilian-Marian},\ and\ \citenamefont
  {Venugopalan}}]{Gelis:2010nm}%
  \BibitemOpen
  \bibfield  {author} {\bibinfo {author} {\bibfnamefont {F.}~\bibnamefont
  {Gelis}}, \bibinfo {author} {\bibfnamefont {E.}~\bibnamefont {Iancu}},
  \bibinfo {author} {\bibfnamefont {J.}~\bibnamefont {Jalilian-Marian}},\ and\
  \bibinfo {author} {\bibfnamefont {R.}~\bibnamefont {Venugopalan}},\
  }\bibfield  {title} {\bibinfo {title} {{The Color Glass Condensate}},\ }\href
  {https://doi.org/10.1146/annurev.nucl.010909.083629} {\bibfield  {journal}
  {\bibinfo  {journal} {Ann. Rev. Nucl. Part. Sci.}\ }\textbf {\bibinfo
  {volume} {60}},\ \bibinfo {pages} {463} (\bibinfo {year} {2010})},\ \Eprint
  {https://arxiv.org/abs/1002.0333} {arXiv:1002.0333 [hep-ph]} \BibitemShut
  {NoStop}%
\bibitem [{\citenamefont {Dokshitzer}(1977)}]{Dokshitzer:1977sg}%
  \BibitemOpen
  \bibfield  {author} {\bibinfo {author} {\bibfnamefont {Y.~L.}\ \bibnamefont
  {Dokshitzer}},\ }\bibfield  {title} {\bibinfo {title} {{Calculation of the
  Structure Functions for Deep Inelastic Scattering and e+ e- Annihilation by
  Perturbation Theory in Quantum Chromodynamics.}},\ }\href@noop {} {\bibfield
  {journal} {\bibinfo  {journal} {Sov. Phys. JETP}\ }\textbf {\bibinfo {volume}
  {46}},\ \bibinfo {pages} {641} (\bibinfo {year} {1977})}\BibitemShut
  {NoStop}%
\bibitem [{\citenamefont {Gribov}\ and\ \citenamefont
  {Lipatov}(1972{\natexlab{a}})}]{Gribov:1972ri}%
  \BibitemOpen
  \bibfield  {author} {\bibinfo {author} {\bibfnamefont {V.~N.}\ \bibnamefont
  {Gribov}}\ and\ \bibinfo {author} {\bibfnamefont {L.~N.}\ \bibnamefont
  {Lipatov}},\ }\bibfield  {title} {\bibinfo {title} {{Deep inelastic e p
  scattering in perturbation theory}},\ }\href@noop {} {\bibfield  {journal}
  {\bibinfo  {journal} {Sov. J. Nucl. Phys.}\ }\textbf {\bibinfo {volume}
  {15}},\ \bibinfo {pages} {438} (\bibinfo {year}
  {1972}{\natexlab{a}})}\BibitemShut {NoStop}%
\bibitem [{\citenamefont {Gribov}\ and\ \citenamefont
  {Lipatov}(1972{\natexlab{b}})}]{Gribov:1972rt}%
  \BibitemOpen
  \bibfield  {author} {\bibinfo {author} {\bibfnamefont {V.~N.}\ \bibnamefont
  {Gribov}}\ and\ \bibinfo {author} {\bibfnamefont {L.~N.}\ \bibnamefont
  {Lipatov}},\ }\bibfield  {title} {\bibinfo {title} {{e+ e- pair annihilation
  and deep inelastic e p scattering in perturbation theory}},\ }\href@noop {}
  {\bibfield  {journal} {\bibinfo  {journal} {Sov. J. Nucl. Phys.}\ }\textbf
  {\bibinfo {volume} {15}},\ \bibinfo {pages} {675} (\bibinfo {year}
  {1972}{\natexlab{b}})}\BibitemShut {NoStop}%
\bibitem [{\citenamefont {Altarelli}\ and\ \citenamefont
  {Parisi}(1977)}]{Altarelli:1977zs}%
  \BibitemOpen
  \bibfield  {author} {\bibinfo {author} {\bibfnamefont {G.}~\bibnamefont
  {Altarelli}}\ and\ \bibinfo {author} {\bibfnamefont {G.}~\bibnamefont
  {Parisi}},\ }\bibfield  {title} {\bibinfo {title} {{Asymptotic Freedom in
  Parton Language}},\ }\href {https://doi.org/10.1016/0550-3213(77)90384-4}
  {\bibfield  {journal} {\bibinfo  {journal} {Nucl. Phys. B}\ }\textbf
  {\bibinfo {volume} {126}},\ \bibinfo {pages} {298} (\bibinfo {year}
  {1977})}\BibitemShut {NoStop}%
\bibitem [{\citenamefont {Balitsky}\ and\ \citenamefont
  {Lipatov}(1978)}]{Balitsky:1978ic}%
  \BibitemOpen
  \bibfield  {author} {\bibinfo {author} {\bibfnamefont {I.~I.}\ \bibnamefont
  {Balitsky}}\ and\ \bibinfo {author} {\bibfnamefont {L.~N.}\ \bibnamefont
  {Lipatov}},\ }\bibfield  {title} {\bibinfo {title} {{The Pomeranchuk
  Singularity in Quantum Chromodynamics}},\ }\href@noop {} {\bibfield
  {journal} {\bibinfo  {journal} {Sov. J. Nucl. Phys.}\ }\textbf {\bibinfo
  {volume} {28}},\ \bibinfo {pages} {822} (\bibinfo {year} {1978})}\BibitemShut
  {NoStop}%
\bibitem [{\citenamefont {Kuraev}\ \emph {et~al.}(1977)\citenamefont {Kuraev},
  \citenamefont {Lipatov},\ and\ \citenamefont {Fadin}}]{Kuraev:1977fs}%
  \BibitemOpen
  \bibfield  {author} {\bibinfo {author} {\bibfnamefont {E.~A.}\ \bibnamefont
  {Kuraev}}, \bibinfo {author} {\bibfnamefont {L.~N.}\ \bibnamefont
  {Lipatov}},\ and\ \bibinfo {author} {\bibfnamefont {V.~S.}\ \bibnamefont
  {Fadin}},\ }\bibfield  {title} {\bibinfo {title} {{The Pomeranchuk
  Singularity in Nonabelian Gauge Theories}},\ }\href@noop {} {\bibfield
  {journal} {\bibinfo  {journal} {Sov. Phys. JETP}\ }\textbf {\bibinfo {volume}
  {45}},\ \bibinfo {pages} {199} (\bibinfo {year} {1977})}\BibitemShut
  {NoStop}%
\bibitem [{\citenamefont {Gribov}\ \emph {et~al.}(1983)\citenamefont {Gribov},
  \citenamefont {Levin},\ and\ \citenamefont {Ryskin}}]{Gribov:1983ivg}%
  \BibitemOpen
  \bibfield  {author} {\bibinfo {author} {\bibfnamefont {L.~V.}\ \bibnamefont
  {Gribov}}, \bibinfo {author} {\bibfnamefont {E.~M.}\ \bibnamefont {Levin}},\
  and\ \bibinfo {author} {\bibfnamefont {M.~G.}\ \bibnamefont {Ryskin}},\
  }\bibfield  {title} {\bibinfo {title} {{Semihard Processes in QCD}},\ }\href
  {https://doi.org/10.1016/0370-1573(83)90022-4} {\bibfield  {journal}
  {\bibinfo  {journal} {Phys. Rept.}\ }\textbf {\bibinfo {volume} {100}},\
  \bibinfo {pages} {1} (\bibinfo {year} {1983})}\BibitemShut {NoStop}%
\bibitem [{\citenamefont {Mueller}\ and\ \citenamefont
  {Qiu}(1986)}]{Mueller:1985wy}%
  \BibitemOpen
  \bibfield  {author} {\bibinfo {author} {\bibfnamefont {A.~H.}\ \bibnamefont
  {Mueller}}\ and\ \bibinfo {author} {\bibfnamefont {J.-w.}\ \bibnamefont
  {Qiu}},\ }\bibfield  {title} {\bibinfo {title} {{Gluon Recombination and
  Shadowing at Small Values of x}},\ }\href
  {https://doi.org/10.1016/0550-3213(86)90164-1} {\bibfield  {journal}
  {\bibinfo  {journal} {Nucl. Phys. B}\ }\textbf {\bibinfo {volume} {268}},\
  \bibinfo {pages} {427} (\bibinfo {year} {1986})}\BibitemShut {NoStop}%
\bibitem [{\citenamefont {Jalilian-Marian}\ \emph
  {et~al.}(1998{\natexlab{a}})\citenamefont {Jalilian-Marian}, \citenamefont
  {Kovner}, \citenamefont {Leonidov},\ and\ \citenamefont
  {Weigert}}]{Jalilian-Marian:1997jhx}%
  \BibitemOpen
  \bibfield  {author} {\bibinfo {author} {\bibfnamefont {J.}~\bibnamefont
  {Jalilian-Marian}}, \bibinfo {author} {\bibfnamefont {A.}~\bibnamefont
  {Kovner}}, \bibinfo {author} {\bibfnamefont {A.}~\bibnamefont {Leonidov}},\
  and\ \bibinfo {author} {\bibfnamefont {H.}~\bibnamefont {Weigert}},\
  }\bibfield  {title} {\bibinfo {title} {{The Wilson renormalization group for
  low x physics: Towards the high density regime}},\ }\href
  {https://doi.org/10.1103/PhysRevD.59.014014} {\bibfield  {journal} {\bibinfo
  {journal} {Phys. Rev. D}\ }\textbf {\bibinfo {volume} {59}},\ \bibinfo
  {pages} {014014} (\bibinfo {year} {1998}{\natexlab{a}})},\ \Eprint
  {https://arxiv.org/abs/hep-ph/9706377} {arXiv:hep-ph/9706377} \BibitemShut
  {NoStop}%
\bibitem [{\citenamefont {Jalilian-Marian}\ \emph
  {et~al.}(1998{\natexlab{b}})\citenamefont {Jalilian-Marian}, \citenamefont
  {Kovner},\ and\ \citenamefont {Weigert}}]{Jalilian-Marian:1997ubg}%
  \BibitemOpen
  \bibfield  {author} {\bibinfo {author} {\bibfnamefont {J.}~\bibnamefont
  {Jalilian-Marian}}, \bibinfo {author} {\bibfnamefont {A.}~\bibnamefont
  {Kovner}},\ and\ \bibinfo {author} {\bibfnamefont {H.}~\bibnamefont
  {Weigert}},\ }\bibfield  {title} {\bibinfo {title} {{The Wilson
  renormalization group for low x physics: Gluon evolution at finite parton
  density}},\ }\href {https://doi.org/10.1103/PhysRevD.59.014015} {\bibfield
  {journal} {\bibinfo  {journal} {Phys. Rev. D}\ }\textbf {\bibinfo {volume}
  {59}},\ \bibinfo {pages} {014015} (\bibinfo {year} {1998}{\natexlab{b}})},\
  \Eprint {https://arxiv.org/abs/hep-ph/9709432} {arXiv:hep-ph/9709432}
  \BibitemShut {NoStop}%
\bibitem [{\citenamefont {Iancu}\ \emph
  {et~al.}(2001{\natexlab{a}})\citenamefont {Iancu}, \citenamefont {Leonidov},\
  and\ \citenamefont {McLerran}}]{Iancu:2000hn}%
  \BibitemOpen
  \bibfield  {author} {\bibinfo {author} {\bibfnamefont {E.}~\bibnamefont
  {Iancu}}, \bibinfo {author} {\bibfnamefont {A.}~\bibnamefont {Leonidov}},\
  and\ \bibinfo {author} {\bibfnamefont {L.~D.}\ \bibnamefont {McLerran}},\
  }\bibfield  {title} {\bibinfo {title} {{Nonlinear gluon evolution in the
  color glass condensate. 1.}},\ }\href
  {https://doi.org/10.1016/S0375-9474(01)00642-X} {\bibfield  {journal}
  {\bibinfo  {journal} {Nucl. Phys. A}\ }\textbf {\bibinfo {volume} {692}},\
  \bibinfo {pages} {583} (\bibinfo {year} {2001}{\natexlab{a}})},\ \Eprint
  {https://arxiv.org/abs/hep-ph/0011241} {arXiv:hep-ph/0011241} \BibitemShut
  {NoStop}%
\bibitem [{\citenamefont {Iancu}\ \emph
  {et~al.}(2001{\natexlab{b}})\citenamefont {Iancu}, \citenamefont {Leonidov},\
  and\ \citenamefont {McLerran}}]{Iancu:2001ad}%
  \BibitemOpen
  \bibfield  {author} {\bibinfo {author} {\bibfnamefont {E.}~\bibnamefont
  {Iancu}}, \bibinfo {author} {\bibfnamefont {A.}~\bibnamefont {Leonidov}},\
  and\ \bibinfo {author} {\bibfnamefont {L.~D.}\ \bibnamefont {McLerran}},\
  }\bibfield  {title} {\bibinfo {title} {{The Renormalization group equation
  for the color glass condensate}},\ }\href
  {https://doi.org/10.1016/S0370-2693(01)00524-X} {\bibfield  {journal}
  {\bibinfo  {journal} {Phys. Lett. B}\ }\textbf {\bibinfo {volume} {510}},\
  \bibinfo {pages} {133} (\bibinfo {year} {2001}{\natexlab{b}})},\ \Eprint
  {https://arxiv.org/abs/hep-ph/0102009} {arXiv:hep-ph/0102009} \BibitemShut
  {NoStop}%
\bibitem [{\citenamefont {Weigert}(2002)}]{Weigert:2000gi}%
  \BibitemOpen
  \bibfield  {author} {\bibinfo {author} {\bibfnamefont {H.}~\bibnamefont
  {Weigert}},\ }\bibfield  {title} {\bibinfo {title} {{Unitarity at small
  Bjorken x}},\ }\href {https://doi.org/10.1016/S0375-9474(01)01668-2}
  {\bibfield  {journal} {\bibinfo  {journal} {Nucl. Phys. A}\ }\textbf
  {\bibinfo {volume} {703}},\ \bibinfo {pages} {823} (\bibinfo {year}
  {2002})},\ \Eprint {https://arxiv.org/abs/hep-ph/0004044}
  {arXiv:hep-ph/0004044} \BibitemShut {NoStop}%
\bibitem [{\citenamefont {Balitsky}(1996)}]{Balitsky:1995ub}%
  \BibitemOpen
  \bibfield  {author} {\bibinfo {author} {\bibfnamefont {I.}~\bibnamefont
  {Balitsky}},\ }\bibfield  {title} {\bibinfo {title} {{Operator expansion for
  high-energy scattering}},\ }\href
  {https://doi.org/10.1016/0550-3213(95)00638-9} {\bibfield  {journal}
  {\bibinfo  {journal} {Nucl. Phys. B}\ }\textbf {\bibinfo {volume} {463}},\
  \bibinfo {pages} {99} (\bibinfo {year} {1996})},\ \Eprint
  {https://arxiv.org/abs/hep-ph/9509348} {arXiv:hep-ph/9509348} \BibitemShut
  {NoStop}%
\bibitem [{\citenamefont {Kovchegov}(1999)}]{Kovchegov:1999yj}%
  \BibitemOpen
  \bibfield  {author} {\bibinfo {author} {\bibfnamefont {Y.~V.}\ \bibnamefont
  {Kovchegov}},\ }\bibfield  {title} {\bibinfo {title} {{Small x F(2) structure
  function of a nucleus including multiple pomeron exchanges}},\ }\href
  {https://doi.org/10.1103/PhysRevD.60.034008} {\bibfield  {journal} {\bibinfo
  {journal} {Phys. Rev. D}\ }\textbf {\bibinfo {volume} {60}},\ \bibinfo
  {pages} {034008} (\bibinfo {year} {1999})},\ \Eprint
  {https://arxiv.org/abs/hep-ph/9901281} {arXiv:hep-ph/9901281} \BibitemShut
  {NoStop}%
\bibitem [{\citenamefont {Kovchegov}(2000)}]{Kovchegov:1999ua}%
  \BibitemOpen
  \bibfield  {author} {\bibinfo {author} {\bibfnamefont {Y.~V.}\ \bibnamefont
  {Kovchegov}},\ }\bibfield  {title} {\bibinfo {title} {{Unitarization of the
  BFKL pomeron on a nucleus}},\ }\href
  {https://doi.org/10.1103/PhysRevD.61.074018} {\bibfield  {journal} {\bibinfo
  {journal} {Phys. Rev. D}\ }\textbf {\bibinfo {volume} {61}},\ \bibinfo
  {pages} {074018} (\bibinfo {year} {2000})},\ \Eprint
  {https://arxiv.org/abs/hep-ph/9905214} {arXiv:hep-ph/9905214} \BibitemShut
  {NoStop}%
\bibitem [{\citenamefont {Enberg}\ \emph {et~al.}(2005)\citenamefont {Enberg},
  \citenamefont {Golec-Biernat},\ and\ \citenamefont {Munier}}]{Enberg:2005cb}%
  \BibitemOpen
  \bibfield  {author} {\bibinfo {author} {\bibfnamefont {R.}~\bibnamefont
  {Enberg}}, \bibinfo {author} {\bibfnamefont {K.~J.}\ \bibnamefont
  {Golec-Biernat}},\ and\ \bibinfo {author} {\bibfnamefont {S.}~\bibnamefont
  {Munier}},\ }\bibfield  {title} {\bibinfo {title} {{The High energy
  asymptotics of scattering processes in QCD}},\ }\href
  {https://doi.org/10.1103/PhysRevD.72.074021} {\bibfield  {journal} {\bibinfo
  {journal} {Phys. Rev. D}\ }\textbf {\bibinfo {volume} {72}},\ \bibinfo
  {pages} {074021} (\bibinfo {year} {2005})},\ \Eprint
  {https://arxiv.org/abs/hep-ph/0505101} {arXiv:hep-ph/0505101} \BibitemShut
  {NoStop}%
\bibitem [{\citenamefont {Munier}\ and\ \citenamefont
  {Peschanski}(2004{\natexlab{a}})}]{Munier:2003sj}%
  \BibitemOpen
  \bibfield  {author} {\bibinfo {author} {\bibfnamefont {S.}~\bibnamefont
  {Munier}}\ and\ \bibinfo {author} {\bibfnamefont {R.~B.}\ \bibnamefont
  {Peschanski}},\ }\bibfield  {title} {\bibinfo {title} {{Traveling wave fronts
  and the transition to saturation}},\ }\href
  {https://doi.org/10.1103/PhysRevD.69.034008} {\bibfield  {journal} {\bibinfo
  {journal} {Phys. Rev. D}\ }\textbf {\bibinfo {volume} {69}},\ \bibinfo
  {pages} {034008} (\bibinfo {year} {2004}{\natexlab{a}})},\ \Eprint
  {https://arxiv.org/abs/hep-ph/0310357} {arXiv:hep-ph/0310357} \BibitemShut
  {NoStop}%
\bibitem [{\citenamefont {Munier}\ and\ \citenamefont
  {Peschanski}(2003)}]{Munier:2003vc}%
  \BibitemOpen
  \bibfield  {author} {\bibinfo {author} {\bibfnamefont {S.}~\bibnamefont
  {Munier}}\ and\ \bibinfo {author} {\bibfnamefont {R.~B.}\ \bibnamefont
  {Peschanski}},\ }\bibfield  {title} {\bibinfo {title} {{Geometric scaling as
  traveling waves}},\ }\href {https://doi.org/10.1103/PhysRevLett.91.232001}
  {\bibfield  {journal} {\bibinfo  {journal} {Phys. Rev. Lett.}\ }\textbf
  {\bibinfo {volume} {91}},\ \bibinfo {pages} {232001} (\bibinfo {year}
  {2003})},\ \Eprint {https://arxiv.org/abs/hep-ph/0309177}
  {arXiv:hep-ph/0309177} \BibitemShut {NoStop}%
\bibitem [{\citenamefont {Munier}\ and\ \citenamefont
  {Peschanski}(2004{\natexlab{b}})}]{Munier:2004xu}%
  \BibitemOpen
  \bibfield  {author} {\bibinfo {author} {\bibfnamefont {S.}~\bibnamefont
  {Munier}}\ and\ \bibinfo {author} {\bibfnamefont {R.~B.}\ \bibnamefont
  {Peschanski}},\ }\bibfield  {title} {\bibinfo {title} {{Universality and tree
  structure of high-energy QCD}},\ }\href
  {https://doi.org/10.1103/PhysRevD.70.077503} {\bibfield  {journal} {\bibinfo
  {journal} {Phys. Rev. D}\ }\textbf {\bibinfo {volume} {70}},\ \bibinfo
  {pages} {077503} (\bibinfo {year} {2004}{\natexlab{b}})},\ \Eprint
  {https://arxiv.org/abs/hep-ph/0401215} {arXiv:hep-ph/0401215} \BibitemShut
  {NoStop}%
\bibitem [{\citenamefont {Iancu}\ \emph {et~al.}(2005)\citenamefont {Iancu},
  \citenamefont {Mueller},\ and\ \citenamefont {Munier}}]{Iancu:2004es}%
  \BibitemOpen
  \bibfield  {author} {\bibinfo {author} {\bibfnamefont {E.}~\bibnamefont
  {Iancu}}, \bibinfo {author} {\bibfnamefont {A.~H.}\ \bibnamefont {Mueller}},\
  and\ \bibinfo {author} {\bibfnamefont {S.}~\bibnamefont {Munier}},\
  }\bibfield  {title} {\bibinfo {title} {{Universal behavior of QCD amplitudes
  at high energy from general tools of statistical physics}},\ }\href
  {https://doi.org/10.1016/j.physletb.2004.12.009} {\bibfield  {journal}
  {\bibinfo  {journal} {Phys. Lett. B}\ }\textbf {\bibinfo {volume} {606}},\
  \bibinfo {pages} {342} (\bibinfo {year} {2005})},\ \Eprint
  {https://arxiv.org/abs/hep-ph/0410018} {arXiv:hep-ph/0410018} \BibitemShut
  {NoStop}%
\bibitem [{\citenamefont {Enberg}(2005)}]{Enberg:2005uv}%
  \BibitemOpen
  \bibfield  {author} {\bibinfo {author} {\bibfnamefont {R.}~\bibnamefont
  {Enberg}},\ }\bibfield  {title} {\bibinfo {title} {{Saturation, traveling
  waves and fluctuations}},\ }\href {https://doi.org/10.1063/1.2122043}
  {\bibfield  {journal} {\bibinfo  {journal} {AIP Conf. Proc.}\ }\textbf
  {\bibinfo {volume} {792}},\ \bibinfo {pages} {307} (\bibinfo {year}
  {2005})},\ \Eprint {https://arxiv.org/abs/hep-ph/0507153}
  {arXiv:hep-ph/0507153} \BibitemShut {NoStop}%
\bibitem [{\citenamefont {Munier}(2015)}]{Munier:2014bba}%
  \BibitemOpen
  \bibfield  {author} {\bibinfo {author} {\bibfnamefont {S.}~\bibnamefont
  {Munier}},\ }\bibfield  {title} {\bibinfo {title} {{Statistical physics in
  QCD evolution towards high energies}},\ }\href
  {https://doi.org/10.1007/s11433-015-5666-7} {\bibfield  {journal} {\bibinfo
  {journal} {Sci. China Phys. Mech. Astron.}\ }\textbf {\bibinfo {volume}
  {58}},\ \bibinfo {pages} {81001} (\bibinfo {year} {2015})},\ \Eprint
  {https://arxiv.org/abs/1410.6478} {arXiv:1410.6478 [hep-ph]} \BibitemShut
  {NoStop}%
\bibitem [{\citenamefont {Mueller}\ and\ \citenamefont
  {Munier}(2018)}]{Mueller:2018zwx}%
  \BibitemOpen
  \bibfield  {author} {\bibinfo {author} {\bibfnamefont {A.~H.}\ \bibnamefont
  {Mueller}}\ and\ \bibinfo {author} {\bibfnamefont {S.}~\bibnamefont
  {Munier}},\ }\bibfield  {title} {\bibinfo {title} {{Diffractive
  Electron-Nucleus Scattering and Ancestry in Branching Random Walks}},\ }\href
  {https://doi.org/10.1103/PhysRevLett.121.082001} {\bibfield  {journal}
  {\bibinfo  {journal} {Phys. Rev. Lett.}\ }\textbf {\bibinfo {volume} {121}},\
  \bibinfo {pages} {082001} (\bibinfo {year} {2018})},\ \Eprint
  {https://arxiv.org/abs/1805.09417} {arXiv:1805.09417 [hep-ph]} \BibitemShut
  {NoStop}%
\bibitem [{\citenamefont {Yang}\ \emph {et~al.}(2020)\citenamefont {Yang},
  \citenamefont {Kou}, \citenamefont {Wang},\ and\ \citenamefont
  {Chen}}]{Yang:2020jmt}%
  \BibitemOpen
  \bibfield  {author} {\bibinfo {author} {\bibfnamefont {Y.}~\bibnamefont
  {Yang}}, \bibinfo {author} {\bibfnamefont {W.}~\bibnamefont {Kou}}, \bibinfo
  {author} {\bibfnamefont {X.}~\bibnamefont {Wang}},\ and\ \bibinfo {author}
  {\bibfnamefont {X.}~\bibnamefont {Chen}},\ }\bibfield  {title} {\bibinfo
  {title} {{Solitary wave solutions of FKPP equation using Homogeneous balance
  method(HB method)}},\ }\href@noop {} {\  (\bibinfo {year} {2020})},\ \Eprint
  {https://arxiv.org/abs/2009.11378} {arXiv:2009.11378 [nlin.PS]} \BibitemShut
  {NoStop}%
\bibitem [{\citenamefont {Wang}\ \emph {et~al.}(2021)\citenamefont {Wang},
  \citenamefont {Yang}, \citenamefont {Kou}, \citenamefont {Wang},\ and\
  \citenamefont {Chen}}]{Wang:2020stj}%
  \BibitemOpen
  \bibfield  {author} {\bibinfo {author} {\bibfnamefont {X.}~\bibnamefont
  {Wang}}, \bibinfo {author} {\bibfnamefont {Y.}~\bibnamefont {Yang}}, \bibinfo
  {author} {\bibfnamefont {W.}~\bibnamefont {Kou}}, \bibinfo {author}
  {\bibfnamefont {R.}~\bibnamefont {Wang}},\ and\ \bibinfo {author}
  {\bibfnamefont {X.}~\bibnamefont {Chen}},\ }\bibfield  {title} {\bibinfo
  {title} {{Analytical solution of Balitsky-Kovchegov equation with homogeneous
  balance method}},\ }\href {https://doi.org/10.1103/PhysRevD.103.056008}
  {\bibfield  {journal} {\bibinfo  {journal} {Phys. Rev. D}\ }\textbf {\bibinfo
  {volume} {103}},\ \bibinfo {pages} {056008} (\bibinfo {year} {2021})},\
  \Eprint {https://arxiv.org/abs/2009.13325} {arXiv:2009.13325 [hep-ph]}
  \BibitemShut {NoStop}%
\bibitem [{\citenamefont {Wang}\ \emph {et~al.}(2022)\citenamefont {Wang},
  \citenamefont {Kou}, \citenamefont {Xie}, \citenamefont {Xie},\ and\
  \citenamefont {Chen}}]{Wang:2022jwh}%
  \BibitemOpen
  \bibfield  {author} {\bibinfo {author} {\bibfnamefont {X.}~\bibnamefont
  {Wang}}, \bibinfo {author} {\bibfnamefont {W.}~\bibnamefont {Kou}}, \bibinfo
  {author} {\bibfnamefont {G.}~\bibnamefont {Xie}}, \bibinfo {author}
  {\bibfnamefont {Y.-P.}\ \bibnamefont {Xie}},\ and\ \bibinfo {author}
  {\bibfnamefont {X.}~\bibnamefont {Chen}},\ }\bibfield  {title} {\bibinfo
  {title} {{Exclusive vector meson production with the analytical solution of
  Balitsky-Kovchegov equation}},\ }\href
  {https://doi.org/10.1088/1674-1137/ac6daa} {\bibfield  {journal} {\bibinfo
  {journal} {Chin. Phys. C}\ }\textbf {\bibinfo {volume} {46}},\ \bibinfo
  {pages} {093101} (\bibinfo {year} {2022})},\ \Eprint
  {https://arxiv.org/abs/2205.02396} {arXiv:2205.02396 [hep-ph]} \BibitemShut
  {NoStop}%
\bibitem [{\citenamefont {Cai}\ \emph {et~al.}(2023)\citenamefont {Cai},
  \citenamefont {Wang},\ and\ \citenamefont {Chen}}]{Cai:2023iza}%
  \BibitemOpen
  \bibfield  {author} {\bibinfo {author} {\bibfnamefont {Y.}~\bibnamefont
  {Cai}}, \bibinfo {author} {\bibfnamefont {X.}~\bibnamefont {Wang}},\ and\
  \bibinfo {author} {\bibfnamefont {X.}~\bibnamefont {Chen}},\ }\bibfield
  {title} {\bibinfo {title} {{Analytic solution of the Balitsky-Kovchegov
  equation with a running coupling constant using the homogeneous balance
  method}},\ }\href {https://doi.org/10.1103/PhysRevD.108.116024} {\bibfield
  {journal} {\bibinfo  {journal} {Phys. Rev. D}\ }\textbf {\bibinfo {volume}
  {108}},\ \bibinfo {pages} {116024} (\bibinfo {year} {2023})},\ \Eprint
  {https://arxiv.org/abs/2311.02672} {arXiv:2311.02672 [hep-ph]} \BibitemShut
  {NoStop}%
\bibitem [{\citenamefont {Balitsky}(2007)}]{Balitsky:2006wa}%
  \BibitemOpen
  \bibfield  {author} {\bibinfo {author} {\bibfnamefont {I.}~\bibnamefont
  {Balitsky}},\ }\bibfield  {title} {\bibinfo {title} {{Quark contribution to
  the small-x evolution of color dipole}},\ }\href
  {https://doi.org/10.1103/PhysRevD.75.014001} {\bibfield  {journal} {\bibinfo
  {journal} {Phys. Rev. D}\ }\textbf {\bibinfo {volume} {75}},\ \bibinfo
  {pages} {014001} (\bibinfo {year} {2007})},\ \Eprint
  {https://arxiv.org/abs/hep-ph/0609105} {arXiv:hep-ph/0609105} \BibitemShut
  {NoStop}%
\bibitem [{\citenamefont {Iancu}\ \emph
  {et~al.}(2002{\natexlab{b}})\citenamefont {Iancu}, \citenamefont {Itakura},\
  and\ \citenamefont {McLerran}}]{Iancu:2002tr}%
  \BibitemOpen
  \bibfield  {author} {\bibinfo {author} {\bibfnamefont {E.}~\bibnamefont
  {Iancu}}, \bibinfo {author} {\bibfnamefont {K.}~\bibnamefont {Itakura}},\
  and\ \bibinfo {author} {\bibfnamefont {L.}~\bibnamefont {McLerran}},\
  }\bibfield  {title} {\bibinfo {title} {{Geometric scaling above the
  saturation scale}},\ }\href {https://doi.org/10.1016/S0375-9474(02)01010-2}
  {\bibfield  {journal} {\bibinfo  {journal} {Nucl. Phys. A}\ }\textbf
  {\bibinfo {volume} {708}},\ \bibinfo {pages} {327} (\bibinfo {year}
  {2002}{\natexlab{b}})},\ \Eprint {https://arxiv.org/abs/hep-ph/0203137}
  {arXiv:hep-ph/0203137} \BibitemShut {NoStop}%
\bibitem [{\citenamefont {Aaron}\ \emph {et~al.}(2010)\citenamefont {Aaron}
  \emph {et~al.}}]{H1:2009pze}%
  \BibitemOpen
  \bibfield  {author} {\bibinfo {author} {\bibfnamefont {F.~D.}\ \bibnamefont
  {Aaron}} \emph {et~al.} (\bibinfo {collaboration} {H1, ZEUS}),\ }\bibfield
  {title} {\bibinfo {title} {{Combined Measurement and QCD Analysis of the
  Inclusive e+- p Scattering Cross Sections at HERA}},\ }\href
  {https://doi.org/10.1007/JHEP01(2010)109} {\bibfield  {journal} {\bibinfo
  {journal} {JHEP}\ }\textbf {\bibinfo {volume} {01}},\ \bibinfo {pages}
  {109}},\ \Eprint {https://arxiv.org/abs/0911.0884} {arXiv:0911.0884 [hep-ex]}
  \BibitemShut {NoStop}%
\bibitem [{\citenamefont {Andreev}\ \emph {et~al.}(2014)\citenamefont {Andreev}
  \emph {et~al.}}]{H1:2013ktq}%
  \BibitemOpen
  \bibfield  {author} {\bibinfo {author} {\bibfnamefont {V.}~\bibnamefont
  {Andreev}} \emph {et~al.} (\bibinfo {collaboration} {H1}),\ }\bibfield
  {title} {\bibinfo {title} {{Measurement of inclusive $e p$ cross sections at
  high $Q^2$ at $\sqrt s =$ 225 and 252 GeV and of the longitudinal proton
  structure function $F_L$ at HERA}},\ }\href
  {https://doi.org/10.1140/epjc/s10052-014-2814-6} {\bibfield  {journal}
  {\bibinfo  {journal} {Eur. Phys. J. C}\ }\textbf {\bibinfo {volume} {74}},\
  \bibinfo {pages} {2814} (\bibinfo {year} {2014})},\ \Eprint
  {https://arxiv.org/abs/1312.4821} {arXiv:1312.4821 [hep-ex]} \BibitemShut
  {NoStop}%
\bibitem [{\citenamefont {Iancu}\ and\ \citenamefont
  {Venugopalan}(2003)}]{Iancu:2003xm}%
  \BibitemOpen
  \bibfield  {author} {\bibinfo {author} {\bibfnamefont {E.}~\bibnamefont
  {Iancu}}\ and\ \bibinfo {author} {\bibfnamefont {R.}~\bibnamefont
  {Venugopalan}},\ }\bibinfo {title} {{The Color glass condensate and
  high-energy scattering in QCD}},\ in\ \href
  {https://doi.org/10.1142/9789812795533_0005} {\emph {\bibinfo {booktitle}
  {{Quark-gluon plasma 4}}}},\ \bibinfo {editor} {edited by\ \bibinfo {editor}
  {\bibfnamefont {R.~C.}\ \bibnamefont {Hwa}}\ and\ \bibinfo {editor}
  {\bibfnamefont {X.-N.}\ \bibnamefont {Wang}}}\ (\bibinfo {year} {2003})\ pp.\
  \bibinfo {pages} {249--3363},\ \Eprint {https://arxiv.org/abs/hep-ph/0303204}
  {arXiv:hep-ph/0303204} \BibitemShut {NoStop}%
\bibitem [{\citenamefont {Kutak}\ and\ \citenamefont
  {Stasto}(2005)}]{Kutak:2004ym}%
  \BibitemOpen
  \bibfield  {author} {\bibinfo {author} {\bibfnamefont {K.}~\bibnamefont
  {Kutak}}\ and\ \bibinfo {author} {\bibfnamefont {A.~M.}\ \bibnamefont
  {Stasto}},\ }\bibfield  {title} {\bibinfo {title} {{Unintegrated gluon
  distribution from modified BK equation}},\ }\href
  {https://doi.org/10.1140/epjc/s2005-02223-0} {\bibfield  {journal} {\bibinfo
  {journal} {Eur. Phys. J. C}\ }\textbf {\bibinfo {volume} {41}},\ \bibinfo
  {pages} {343} (\bibinfo {year} {2005})},\ \Eprint
  {https://arxiv.org/abs/hep-ph/0408117} {arXiv:hep-ph/0408117} \BibitemShut
  {NoStop}%
\bibitem [{\citenamefont {Kharzeev}\ and\ \citenamefont
  {Tuchin}(2005)}]{Kharzeev:2005iz}%
  \BibitemOpen
  \bibfield  {author} {\bibinfo {author} {\bibfnamefont {D.}~\bibnamefont
  {Kharzeev}}\ and\ \bibinfo {author} {\bibfnamefont {K.}~\bibnamefont
  {Tuchin}},\ }\bibfield  {title} {\bibinfo {title} {{From color glass
  condensate to quark gluon plasma through the event horizon}},\ }\href
  {https://doi.org/10.1016/j.nuclphysa.2005.03.001} {\bibfield  {journal}
  {\bibinfo  {journal} {Nucl. Phys. A}\ }\textbf {\bibinfo {volume} {753}},\
  \bibinfo {pages} {316} (\bibinfo {year} {2005})},\ \Eprint
  {https://arxiv.org/abs/hep-ph/0501234} {arXiv:hep-ph/0501234} \BibitemShut
  {NoStop}%
\bibitem [{\citenamefont {Castorina}\ \emph {et~al.}(2007)\citenamefont
  {Castorina}, \citenamefont {Kharzeev},\ and\ \citenamefont
  {Satz}}]{Castorina:2007eb}%
  \BibitemOpen
  \bibfield  {author} {\bibinfo {author} {\bibfnamefont {P.}~\bibnamefont
  {Castorina}}, \bibinfo {author} {\bibfnamefont {D.}~\bibnamefont
  {Kharzeev}},\ and\ \bibinfo {author} {\bibfnamefont {H.}~\bibnamefont
  {Satz}},\ }\bibfield  {title} {\bibinfo {title} {{Thermal Hadronization and
  Hawking-Unruh Radiation in QCD}},\ }\href
  {https://doi.org/10.1140/epjc/s10052-007-0368-6} {\bibfield  {journal}
  {\bibinfo  {journal} {Eur. Phys. J. C}\ }\textbf {\bibinfo {volume} {52}},\
  \bibinfo {pages} {187} (\bibinfo {year} {2007})},\ \Eprint
  {https://arxiv.org/abs/0704.1426} {arXiv:0704.1426 [hep-ph]} \BibitemShut
  {NoStop}%
\bibitem [{\citenamefont {Unruh}(1976)}]{Unruh:1976db}%
  \BibitemOpen
  \bibfield  {author} {\bibinfo {author} {\bibfnamefont {W.~G.}\ \bibnamefont
  {Unruh}},\ }\bibfield  {title} {\bibinfo {title} {{Notes on black hole
  evaporation}},\ }\href {https://doi.org/10.1103/PhysRevD.14.870} {\bibfield
  {journal} {\bibinfo  {journal} {Phys. Rev. D}\ }\textbf {\bibinfo {volume}
  {14}},\ \bibinfo {pages} {870} (\bibinfo {year} {1976})}\BibitemShut
  {NoStop}%
\bibitem [{\citenamefont {Wong}(1970)}]{Wong:1970fu}%
  \BibitemOpen
  \bibfield  {author} {\bibinfo {author} {\bibfnamefont {S.~K.}\ \bibnamefont
  {Wong}},\ }\bibfield  {title} {\bibinfo {title} {{Field and particle
  equations for the classical Yang-Mills field and particles with isotopic
  spin}},\ }\href {https://doi.org/10.1007/BF02892134} {\bibfield  {journal}
  {\bibinfo  {journal} {Nuovo Cim. A}\ }\textbf {\bibinfo {volume} {65}},\
  \bibinfo {pages} {689} (\bibinfo {year} {1970})}\BibitemShut {NoStop}%
\bibitem [{\citenamefont {Mathieu}(2010)}]{Mathieu:2010nbc}%
  \BibitemOpen
  \bibfield  {author} {\bibinfo {author} {\bibfnamefont {V.}~\bibnamefont
  {Mathieu}},\ }\bibfield  {title} {\bibinfo {title} {{Gluon Mass, Glueballs
  and Gluonic Mesons}},\ }\href {https://doi.org/10.22323/1.117.0002}
  {\bibfield  {journal} {\bibinfo  {journal} {PoS}\ }\textbf {\bibinfo {volume}
  {FACESQCD}},\ \bibinfo {pages} {002} (\bibinfo {year} {2010})},\ \Eprint
  {https://arxiv.org/abs/1102.3875} {arXiv:1102.3875 [hep-ph]} \BibitemShut
  {NoStop}%
\bibitem [{\citenamefont {Mathieu}(2011)}]{Mathieu:2011zzb}%
  \BibitemOpen
  \bibfield  {author} {\bibinfo {author} {\bibfnamefont {V.}~\bibnamefont
  {Mathieu}},\ }\bibfield  {title} {\bibinfo {title} {{Glueball
  spectroscopy}},\ }\href {https://doi.org/10.5506/APhysPolBSupp.4.677}
  {\bibfield  {journal} {\bibinfo  {journal} {Acta Phys. Polon. Supp.}\
  }\textbf {\bibinfo {volume} {4}},\ \bibinfo {pages} {677} (\bibinfo {year}
  {2011})}\BibitemShut {NoStop}%
\bibitem [{\citenamefont {Golec-Biernat}\ and\ \citenamefont
  {Wusthoff}(1998)}]{Golec-Biernat:1998zce}%
  \BibitemOpen
  \bibfield  {author} {\bibinfo {author} {\bibfnamefont {K.~J.}\ \bibnamefont
  {Golec-Biernat}}\ and\ \bibinfo {author} {\bibfnamefont {M.}~\bibnamefont
  {Wusthoff}},\ }\bibfield  {title} {\bibinfo {title} {{Saturation effects in
  deep inelastic scattering at low Q**2 and its implications on diffraction}},\
  }\href {https://doi.org/10.1103/PhysRevD.59.014017} {\bibfield  {journal}
  {\bibinfo  {journal} {Phys. Rev. D}\ }\textbf {\bibinfo {volume} {59}},\
  \bibinfo {pages} {014017} (\bibinfo {year} {1998})},\ \Eprint
  {https://arxiv.org/abs/hep-ph/9807513} {arXiv:hep-ph/9807513} \BibitemShut
  {NoStop}%
\bibitem [{\citenamefont {Kharzeev}\ and\ \citenamefont
  {Levin}(2017)}]{Kharzeev:2017qzs}%
  \BibitemOpen
  \bibfield  {author} {\bibinfo {author} {\bibfnamefont {D.~E.}\ \bibnamefont
  {Kharzeev}}\ and\ \bibinfo {author} {\bibfnamefont {E.~M.}\ \bibnamefont
  {Levin}},\ }\bibfield  {title} {\bibinfo {title} {{Deep inelastic scattering
  as a probe of entanglement}},\ }\href
  {https://doi.org/10.1103/PhysRevD.95.114008} {\bibfield  {journal} {\bibinfo
  {journal} {Phys. Rev. D}\ }\textbf {\bibinfo {volume} {95}},\ \bibinfo
  {pages} {114008} (\bibinfo {year} {2017})},\ \Eprint
  {https://arxiv.org/abs/1702.03489} {arXiv:1702.03489 [hep-ph]} \BibitemShut
  {NoStop}%
\bibitem [{\citenamefont {Hentschinski}\ and\ \citenamefont
  {Kutak}(2022)}]{Hentschinski:2021aux}%
  \BibitemOpen
  \bibfield  {author} {\bibinfo {author} {\bibfnamefont {M.}~\bibnamefont
  {Hentschinski}}\ and\ \bibinfo {author} {\bibfnamefont {K.}~\bibnamefont
  {Kutak}},\ }\bibfield  {title} {\bibinfo {title} {{Evidence for the maximally
  entangled low x proton in Deep Inelastic Scattering from H1 data}},\ }\href
  {https://doi.org/10.1140/epjc/s10052-022-10056-y} {\bibfield  {journal}
  {\bibinfo  {journal} {Eur. Phys. J. C}\ }\textbf {\bibinfo {volume} {82}},\
  \bibinfo {pages} {111} (\bibinfo {year} {2022})},\ \bibinfo {note} {[Erratum:
  Eur.Phys.J.C 83, 1147 (2023)]},\ \Eprint {https://arxiv.org/abs/2110.06156}
  {arXiv:2110.06156 [hep-ph]} \BibitemShut {NoStop}%
\bibitem [{\citenamefont {Hentschinski}\ \emph {et~al.}(2022)\citenamefont
  {Hentschinski}, \citenamefont {Kutak},\ and\ \citenamefont
  {Straka}}]{Hentschinski:2022rsa}%
  \BibitemOpen
  \bibfield  {author} {\bibinfo {author} {\bibfnamefont {M.}~\bibnamefont
  {Hentschinski}}, \bibinfo {author} {\bibfnamefont {K.}~\bibnamefont
  {Kutak}},\ and\ \bibinfo {author} {\bibfnamefont {R.}~\bibnamefont
  {Straka}},\ }\bibfield  {title} {\bibinfo {title} {{Maximally entangled
  proton and charged hadron multiplicity in Deep Inelastic Scattering}},\
  }\href {https://doi.org/10.1140/epjc/s10052-022-11122-1} {\bibfield
  {journal} {\bibinfo  {journal} {Eur. Phys. J. C}\ }\textbf {\bibinfo {volume}
  {82}},\ \bibinfo {pages} {1147} (\bibinfo {year} {2022})},\ \Eprint
  {https://arxiv.org/abs/2207.09430} {arXiv:2207.09430 [hep-ph]} \BibitemShut
  {NoStop}%
\bibitem [{\citenamefont {Hentschinski}\ \emph {et~al.}(2023)\citenamefont
  {Hentschinski}, \citenamefont {Kharzeev}, \citenamefont {Kutak},\ and\
  \citenamefont {Tu}}]{Hentschinski:2023izh}%
  \BibitemOpen
  \bibfield  {author} {\bibinfo {author} {\bibfnamefont {M.}~\bibnamefont
  {Hentschinski}}, \bibinfo {author} {\bibfnamefont {D.~E.}\ \bibnamefont
  {Kharzeev}}, \bibinfo {author} {\bibfnamefont {K.}~\bibnamefont {Kutak}},\
  and\ \bibinfo {author} {\bibfnamefont {Z.}~\bibnamefont {Tu}},\ }\bibfield
  {title} {\bibinfo {title} {{Probing the Onset of Maximal Entanglement inside
  the Proton in Diffractive Deep Inelastic Scattering}},\ }\href
  {https://doi.org/10.1103/PhysRevLett.131.241901} {\bibfield  {journal}
  {\bibinfo  {journal} {Phys. Rev. Lett.}\ }\textbf {\bibinfo {volume} {131}},\
  \bibinfo {pages} {241901} (\bibinfo {year} {2023})},\ \Eprint
  {https://arxiv.org/abs/2305.03069} {arXiv:2305.03069 [hep-ph]} \BibitemShut
  {NoStop}%
\bibitem [{\citenamefont {Andreev}\ \emph {et~al.}(2021)\citenamefont {Andreev}
  \emph {et~al.}}]{H1:2020zpd}%
  \BibitemOpen
  \bibfield  {author} {\bibinfo {author} {\bibfnamefont {V.}~\bibnamefont
  {Andreev}} \emph {et~al.} (\bibinfo {collaboration} {H1}),\ }\bibfield
  {title} {\bibinfo {title} {{Measurement of charged particle multiplicity
  distributions in DIS at HERA and its implication to entanglement entropy of
  partons}},\ }\href {https://doi.org/10.1140/epjc/s10052-021-08896-1}
  {\bibfield  {journal} {\bibinfo  {journal} {Eur. Phys. J. C}\ }\textbf
  {\bibinfo {volume} {81}},\ \bibinfo {pages} {212} (\bibinfo {year} {2021})},\
  \Eprint {https://arxiv.org/abs/2011.01812} {arXiv:2011.01812 [hep-ex]}
  \BibitemShut {NoStop}%
\bibitem [{\citenamefont {Tu}\ \emph {et~al.}(2020)\citenamefont {Tu},
  \citenamefont {Kharzeev},\ and\ \citenamefont {Ullrich}}]{Tu:2019ouv}%
  \BibitemOpen
  \bibfield  {author} {\bibinfo {author} {\bibfnamefont {Z.}~\bibnamefont
  {Tu}}, \bibinfo {author} {\bibfnamefont {D.~E.}\ \bibnamefont {Kharzeev}},\
  and\ \bibinfo {author} {\bibfnamefont {T.}~\bibnamefont {Ullrich}},\
  }\bibfield  {title} {\bibinfo {title} {{Einstein-Podolsky-Rosen Paradox and
  Quantum Entanglement at Subnucleonic Scales}},\ }\href
  {https://doi.org/10.1103/PhysRevLett.124.062001} {\bibfield  {journal}
  {\bibinfo  {journal} {Phys. Rev. Lett.}\ }\textbf {\bibinfo {volume} {124}},\
  \bibinfo {pages} {062001} (\bibinfo {year} {2020})},\ \Eprint
  {https://arxiv.org/abs/1904.11974} {arXiv:1904.11974 [hep-ph]} \BibitemShut
  {NoStop}%
\bibitem [{\citenamefont {Glauber}(1955)}]{Glauber:1955qq}%
  \BibitemOpen
  \bibfield  {author} {\bibinfo {author} {\bibfnamefont {R.~J.}\ \bibnamefont
  {Glauber}},\ }\bibfield  {title} {\bibinfo {title} {{Cross-sections in
  deuterium at high-energies}},\ }\href
  {https://doi.org/10.1103/PhysRev.100.242} {\bibfield  {journal} {\bibinfo
  {journal} {Phys. Rev.}\ }\textbf {\bibinfo {volume} {100}},\ \bibinfo {pages}
  {242} (\bibinfo {year} {1955})}\BibitemShut {NoStop}%
\bibitem [{\citenamefont {Gribov}(1970)}]{Gribov:1968gs}%
  \BibitemOpen
  \bibfield  {author} {\bibinfo {author} {\bibfnamefont {V.~N.}\ \bibnamefont
  {Gribov}},\ }\bibfield  {title} {\bibinfo {title} {{Interaction of Gamma
  Quanta and Electrons with Nuclei at High Energies}},\ }\href@noop {}
  {\bibfield  {journal} {\bibinfo  {journal} {Sov. Phys. JETP}\ }\textbf
  {\bibinfo {volume} {30}},\ \bibinfo {pages} {709} (\bibinfo {year}
  {1970})}\BibitemShut {NoStop}%
\bibitem [{\citenamefont {Gribov}(1969)}]{Gribov:1968jf}%
  \BibitemOpen
  \bibfield  {author} {\bibinfo {author} {\bibfnamefont {V.~N.}\ \bibnamefont
  {Gribov}},\ }\bibfield  {title} {\bibinfo {title} {{Glauber Corrections and
  the Interaction between High-energy Hadrons and Nuclei}},\ }\href@noop {}
  {\bibfield  {journal} {\bibinfo  {journal} {Sov. Phys. JETP}\ }\textbf
  {\bibinfo {volume} {29}},\ \bibinfo {pages} {483} (\bibinfo {year}
  {1969})}\BibitemShut {NoStop}%
\bibitem [{\citenamefont {Mueller}(1990)}]{Mueller:1989st}%
  \BibitemOpen
  \bibfield  {author} {\bibinfo {author} {\bibfnamefont {A.~H.}\ \bibnamefont
  {Mueller}},\ }\bibfield  {title} {\bibinfo {title} {{Small x Behavior and
  Parton Saturation: A QCD Model}},\ }\href
  {https://doi.org/10.1016/0550-3213(90)90173-B} {\bibfield  {journal}
  {\bibinfo  {journal} {Nucl. Phys. B}\ }\textbf {\bibinfo {volume} {335}},\
  \bibinfo {pages} {115} (\bibinfo {year} {1990})}\BibitemShut {NoStop}%
\bibitem [{\citenamefont {Mueller}(1999)}]{Mueller:1999wm}%
  \BibitemOpen
  \bibfield  {author} {\bibinfo {author} {\bibfnamefont {A.~H.}\ \bibnamefont
  {Mueller}},\ }\bibfield  {title} {\bibinfo {title} {{Parton saturation at
  small x and in large nuclei}},\ }\href
  {https://doi.org/10.1016/S0550-3213(99)00394-6} {\bibfield  {journal}
  {\bibinfo  {journal} {Nucl. Phys. B}\ }\textbf {\bibinfo {volume} {558}},\
  \bibinfo {pages} {285} (\bibinfo {year} {1999})},\ \Eprint
  {https://arxiv.org/abs/hep-ph/9904404} {arXiv:hep-ph/9904404} \BibitemShut
  {NoStop}%
\bibitem [{\citenamefont {Kovchegov}(2014)}]{Kovchegov:2014kua}%
  \BibitemOpen
  \bibfield  {author} {\bibinfo {author} {\bibfnamefont {Y.~V.}\ \bibnamefont
  {Kovchegov}},\ }\bibfield  {title} {\bibinfo {title} {{Brief Review of
  Saturation Physics}},\ }\href {https://doi.org/10.5506/APhysPolB.45.2241}
  {\bibfield  {journal} {\bibinfo  {journal} {Acta Phys. Polon. B}\ }\textbf
  {\bibinfo {volume} {45}},\ \bibinfo {pages} {2241} (\bibinfo {year}
  {2014})},\ \Eprint {https://arxiv.org/abs/1410.7722} {arXiv:1410.7722
  [hep-ph]} \BibitemShut {NoStop}%
\bibitem [{\citenamefont {Lappi}\ and\ \citenamefont
  {M{\"a}ntysaari}(2013)}]{Lappi:2013zma}%
  \BibitemOpen
  \bibfield  {author} {\bibinfo {author} {\bibfnamefont {T.}~\bibnamefont
  {Lappi}}\ and\ \bibinfo {author} {\bibfnamefont {H.}~\bibnamefont
  {M{\"a}ntysaari}},\ }\bibfield  {title} {\bibinfo {title} {{Single inclusive
  particle production at high energy from HERA data to proton-nucleus
  collisions}},\ }\href {https://doi.org/10.1103/PhysRevD.88.114020} {\bibfield
   {journal} {\bibinfo  {journal} {Phys. Rev. D}\ }\textbf {\bibinfo {volume}
  {88}},\ \bibinfo {pages} {114020} (\bibinfo {year} {2013})},\ \Eprint
  {https://arxiv.org/abs/1309.6963} {arXiv:1309.6963 [hep-ph]} \BibitemShut
  {NoStop}%
\bibitem [{\citenamefont {Deganutti}\ \emph {et~al.}(2024)\citenamefont
  {Deganutti}, \citenamefont {Royon},\ and\ \citenamefont
  {Schlichting}}]{Deganutti:2023qct}%
  \BibitemOpen
  \bibfield  {author} {\bibinfo {author} {\bibfnamefont {F.}~\bibnamefont
  {Deganutti}}, \bibinfo {author} {\bibfnamefont {C.}~\bibnamefont {Royon}},\
  and\ \bibinfo {author} {\bibfnamefont {S.}~\bibnamefont {Schlichting}},\
  }\bibfield  {title} {\bibinfo {title} {{Forward dijet production at the LHC
  within an impact parameter dependent TMD approach}},\ }\href
  {https://doi.org/10.1007/JHEP01(2024)159} {\bibfield  {journal} {\bibinfo
  {journal} {JHEP}\ }\textbf {\bibinfo {volume} {01}},\ \bibinfo {pages}
  {159}},\ \Eprint {https://arxiv.org/abs/2311.01965} {arXiv:2311.01965
  [hep-ph]} \BibitemShut {NoStop}%
\bibitem [{\citenamefont {Armesto}(2003)}]{Armesto:2003pq}%
  \BibitemOpen
  \bibfield  {author} {\bibinfo {author} {\bibfnamefont {N.}~\bibnamefont
  {Armesto}},\ }\bibfield  {title} {\bibinfo {title} {{Nuclear structure
  functions at small x in multiple scattering approaches}},\ }in\ \href@noop {}
  {\emph {\bibinfo {booktitle} {{38th Rencontres de Moriond on QCD and
  High-Energy Hadronic Interactions}}}}\ (\bibinfo {year} {2003})\ \Eprint
  {https://arxiv.org/abs/hep-ph/0305057} {arXiv:hep-ph/0305057} \BibitemShut
  {NoStop}%
\bibitem [{\citenamefont {Woods}\ and\ \citenamefont
  {Saxon}(1954)}]{Woods:1954zz}%
  \BibitemOpen
  \bibfield  {author} {\bibinfo {author} {\bibfnamefont {R.~D.}\ \bibnamefont
  {Woods}}\ and\ \bibinfo {author} {\bibfnamefont {D.~S.}\ \bibnamefont
  {Saxon}},\ }\bibfield  {title} {\bibinfo {title} {{Diffuse Surface Optical
  Model for Nucleon-Nuclei Scattering}},\ }\href
  {https://doi.org/10.1103/PhysRev.95.577} {\bibfield  {journal} {\bibinfo
  {journal} {Phys. Rev.}\ }\textbf {\bibinfo {volume} {95}},\ \bibinfo {pages}
  {577} (\bibinfo {year} {1954})}\BibitemShut {NoStop}%
\bibitem [{\citenamefont {Kovchegov}\ and\ \citenamefont
  {Levin}(2013)}]{Kovchegov:2012mbw}%
  \BibitemOpen
  \bibfield  {author} {\bibinfo {author} {\bibfnamefont {Y.~V.}\ \bibnamefont
  {Kovchegov}}\ and\ \bibinfo {author} {\bibfnamefont {E.}~\bibnamefont
  {Levin}},\ }\href {https://doi.org/10.1017/9781009291446} {\emph {\bibinfo
  {title} {{Quantum Chromodynamics at High Energy}}}},\ Vol.~\bibinfo {volume}
  {33}\ (\bibinfo  {publisher} {Oxford University Press},\ \bibinfo {year}
  {2013})\BibitemShut {NoStop}%
\bibitem [{\citenamefont {M{\"a}ntysaari}\ \emph {et~al.}(2025)\citenamefont
  {M{\"a}ntysaari}, \citenamefont {Penttala}, \citenamefont {Salazar},\ and\
  \citenamefont {Schenke}}]{Mantysaari:2024zxq}%
  \BibitemOpen
  \bibfield  {author} {\bibinfo {author} {\bibfnamefont {H.}~\bibnamefont
  {M{\"a}ntysaari}}, \bibinfo {author} {\bibfnamefont {J.}~\bibnamefont
  {Penttala}}, \bibinfo {author} {\bibfnamefont {F.}~\bibnamefont {Salazar}},\
  and\ \bibinfo {author} {\bibfnamefont {B.}~\bibnamefont {Schenke}},\
  }\bibfield  {title} {\bibinfo {title} {{Finite-size effects on small-x
  evolution and saturation in proton and nuclear targets}},\ }\href
  {https://doi.org/10.1103/PhysRevD.111.054033} {\bibfield  {journal} {\bibinfo
   {journal} {Phys. Rev. D}\ }\textbf {\bibinfo {volume} {111}},\ \bibinfo
  {pages} {054033} (\bibinfo {year} {2025})},\ \Eprint
  {https://arxiv.org/abs/2411.13533} {arXiv:2411.13533 [hep-ph]} \BibitemShut
  {NoStop}%
\bibitem [{\citenamefont {Iancu}\ \emph {et~al.}(2015)\citenamefont {Iancu},
  \citenamefont {Madrigal}, \citenamefont {Mueller}, \citenamefont {Soyez},\
  and\ \citenamefont {Triantafyllopoulos}}]{Iancu:2015vea}%
  \BibitemOpen
  \bibfield  {author} {\bibinfo {author} {\bibfnamefont {E.}~\bibnamefont
  {Iancu}}, \bibinfo {author} {\bibfnamefont {J.~D.}\ \bibnamefont {Madrigal}},
  \bibinfo {author} {\bibfnamefont {A.~H.}\ \bibnamefont {Mueller}}, \bibinfo
  {author} {\bibfnamefont {G.}~\bibnamefont {Soyez}},\ and\ \bibinfo {author}
  {\bibfnamefont {D.~N.}\ \bibnamefont {Triantafyllopoulos}},\ }\bibfield
  {title} {\bibinfo {title} {{Resumming double logarithms in the QCD evolution
  of color dipoles}},\ }\href {https://doi.org/10.1016/j.physletb.2015.03.068}
  {\bibfield  {journal} {\bibinfo  {journal} {Phys. Lett. B}\ }\textbf
  {\bibinfo {volume} {744}},\ \bibinfo {pages} {293} (\bibinfo {year}
  {2015})},\ \Eprint {https://arxiv.org/abs/1502.05642} {arXiv:1502.05642
  [hep-ph]} \BibitemShut {NoStop}%
\bibitem [{\citenamefont {Duclou{\'e}}\ \emph {et~al.}(2019)\citenamefont
  {Duclou{\'e}}, \citenamefont {Iancu}, \citenamefont {Mueller}, \citenamefont
  {Soyez},\ and\ \citenamefont {Triantafyllopoulos}}]{Ducloue:2019ezk}%
  \BibitemOpen
  \bibfield  {author} {\bibinfo {author} {\bibfnamefont {B.}~\bibnamefont
  {Duclou{\'e}}}, \bibinfo {author} {\bibfnamefont {E.}~\bibnamefont {Iancu}},
  \bibinfo {author} {\bibfnamefont {A.~H.}\ \bibnamefont {Mueller}}, \bibinfo
  {author} {\bibfnamefont {G.}~\bibnamefont {Soyez}},\ and\ \bibinfo {author}
  {\bibfnamefont {D.~N.}\ \bibnamefont {Triantafyllopoulos}},\ }\bibfield
  {title} {\bibinfo {title} {{Non-linear evolution in QCD at high-energy beyond
  leading order}},\ }\href {https://doi.org/10.1007/JHEP04(2019)081} {\bibfield
   {journal} {\bibinfo  {journal} {JHEP}\ }\textbf {\bibinfo {volume} {04}},\
  \bibinfo {pages} {081}},\ \Eprint {https://arxiv.org/abs/1902.06637}
  {arXiv:1902.06637 [hep-ph]} \BibitemShut {NoStop}%
\bibitem [{\citenamefont {Koba}\ \emph {et~al.}(1972)\citenamefont {Koba},
  \citenamefont {Nielsen},\ and\ \citenamefont {Olesen}}]{Koba:1972ng}%
  \BibitemOpen
  \bibfield  {author} {\bibinfo {author} {\bibfnamefont {Z.}~\bibnamefont
  {Koba}}, \bibinfo {author} {\bibfnamefont {H.~B.}\ \bibnamefont {Nielsen}},\
  and\ \bibinfo {author} {\bibfnamefont {P.}~\bibnamefont {Olesen}},\
  }\bibfield  {title} {\bibinfo {title} {{Scaling of multiplicity distributions
  in high energy hadron collisions}},\ }\href
  {https://doi.org/10.1016/0550-3213(72)90551-2} {\bibfield  {journal}
  {\bibinfo  {journal} {Nucl. Phys. B}\ }\textbf {\bibinfo {volume} {40}},\
  \bibinfo {pages} {317} (\bibinfo {year} {1972})}\BibitemShut {NoStop}%
\bibitem [{\citenamefont {Altinoluk}\ \emph {et~al.}(2015)\citenamefont
  {Altinoluk}, \citenamefont {Armesto}, \citenamefont {Beuf}, \citenamefont
  {Kovner},\ and\ \citenamefont {Lublinsky}}]{Altinoluk:2015uaa}%
  \BibitemOpen
  \bibfield  {author} {\bibinfo {author} {\bibfnamefont {T.}~\bibnamefont
  {Altinoluk}}, \bibinfo {author} {\bibfnamefont {N.}~\bibnamefont {Armesto}},
  \bibinfo {author} {\bibfnamefont {G.}~\bibnamefont {Beuf}}, \bibinfo {author}
  {\bibfnamefont {A.}~\bibnamefont {Kovner}},\ and\ \bibinfo {author}
  {\bibfnamefont {M.}~\bibnamefont {Lublinsky}},\ }\bibfield  {title} {\bibinfo
  {title} {{Bose enhancement and the ridge}},\ }\href
  {https://doi.org/10.1016/j.physletb.2015.10.072} {\bibfield  {journal}
  {\bibinfo  {journal} {Phys. Lett. B}\ }\textbf {\bibinfo {volume} {751}},\
  \bibinfo {pages} {448} (\bibinfo {year} {2015})},\ \Eprint
  {https://arxiv.org/abs/1503.07126} {arXiv:1503.07126 [hep-ph]} \BibitemShut
  {NoStop}%
\bibitem [{\citenamefont {Accardi}\ \emph {et~al.}(2016)\citenamefont {Accardi}
  \emph {et~al.}}]{Accardi:2012qut}%
  \BibitemOpen
  \bibfield  {author} {\bibinfo {author} {\bibfnamefont {A.}~\bibnamefont
  {Accardi}} \emph {et~al.},\ }\bibfield  {title} {\bibinfo {title} {{Electron
  Ion Collider: The Next QCD Frontier}: {Understanding the glue that binds us
  all}},\ }\href {https://doi.org/10.1140/epja/i2016-16268-9} {\bibfield
  {journal} {\bibinfo  {journal} {Eur. Phys. J. A}\ }\textbf {\bibinfo {volume}
  {52}},\ \bibinfo {pages} {268} (\bibinfo {year} {2016})},\ \Eprint
  {https://arxiv.org/abs/1212.1701} {arXiv:1212.1701 [nucl-ex]} \BibitemShut
  {NoStop}%
\bibitem [{\citenamefont {Aschenauer}\ \emph {et~al.}(2019)\citenamefont
  {Aschenauer}, \citenamefont {Fazio}, \citenamefont {Lee}, \citenamefont
  {Mantysaari}, \citenamefont {Page}, \citenamefont {Schenke}, \citenamefont
  {Ullrich}, \citenamefont {Venugopalan},\ and\ \citenamefont
  {Zurita}}]{Aschenauer:2017jsk}%
  \BibitemOpen
  \bibfield  {author} {\bibinfo {author} {\bibfnamefont {E.~C.}\ \bibnamefont
  {Aschenauer}}, \bibinfo {author} {\bibfnamefont {S.}~\bibnamefont {Fazio}},
  \bibinfo {author} {\bibfnamefont {J.~H.}\ \bibnamefont {Lee}}, \bibinfo
  {author} {\bibfnamefont {H.}~\bibnamefont {Mantysaari}}, \bibinfo {author}
  {\bibfnamefont {B.~S.}\ \bibnamefont {Page}}, \bibinfo {author}
  {\bibfnamefont {B.}~\bibnamefont {Schenke}}, \bibinfo {author} {\bibfnamefont
  {T.}~\bibnamefont {Ullrich}}, \bibinfo {author} {\bibfnamefont
  {R.}~\bibnamefont {Venugopalan}},\ and\ \bibinfo {author} {\bibfnamefont
  {P.}~\bibnamefont {Zurita}},\ }\bibfield  {title} {\bibinfo {title} {{The
  electron{\textendash}ion collider: assessing the energy dependence of key
  measurements}},\ }\href {https://doi.org/10.1088/1361-6633/aaf216} {\bibfield
   {journal} {\bibinfo  {journal} {Rept. Prog. Phys.}\ }\textbf {\bibinfo
  {volume} {82}},\ \bibinfo {pages} {024301} (\bibinfo {year} {2019})},\
  \Eprint {https://arxiv.org/abs/1708.01527} {arXiv:1708.01527 [nucl-ex]}
  \BibitemShut {NoStop}%
\bibitem [{\citenamefont {Abdul~Khalek}\ \emph {et~al.}(2022)\citenamefont
  {Abdul~Khalek} \emph {et~al.}}]{AbdulKhalek:2021gbh}%
  \BibitemOpen
  \bibfield  {author} {\bibinfo {author} {\bibfnamefont {R.}~\bibnamefont
  {Abdul~Khalek}} \emph {et~al.},\ }\bibfield  {title} {\bibinfo {title}
  {{Science Requirements and Detector Concepts for the Electron-Ion Collider}:
  {EIC Yellow Report}},\ }\href
  {https://doi.org/10.1016/j.nuclphysa.2022.122447} {\bibfield  {journal}
  {\bibinfo  {journal} {Nucl. Phys. A}\ }\textbf {\bibinfo {volume} {1026}},\
  \bibinfo {pages} {122447} (\bibinfo {year} {2022})},\ \Eprint
  {https://arxiv.org/abs/2103.05419} {arXiv:2103.05419 [physics.ins-det]}
  \BibitemShut {NoStop}%
\bibitem [{\citenamefont {Morreale}\ and\ \citenamefont
  {Salazar}(2021)}]{Morreale:2021pnn}%
  \BibitemOpen
  \bibfield  {author} {\bibinfo {author} {\bibfnamefont {A.}~\bibnamefont
  {Morreale}}\ and\ \bibinfo {author} {\bibfnamefont {F.}~\bibnamefont
  {Salazar}},\ }\bibfield  {title} {\bibinfo {title} {{Mining for Gluon
  Saturation at Colliders}},\ }\href {https://doi.org/10.3390/universe7080312}
  {\bibfield  {journal} {\bibinfo  {journal} {Universe}\ }\textbf {\bibinfo
  {volume} {7}},\ \bibinfo {pages} {312} (\bibinfo {year} {2021})},\ \Eprint
  {https://arxiv.org/abs/2108.08254} {arXiv:2108.08254 [hep-ph]} \BibitemShut
  {NoStop}%
\bibitem [{\citenamefont {Frankfurt}\ \emph {et~al.}(2005)\citenamefont
  {Frankfurt}, \citenamefont {Strikman},\ and\ \citenamefont
  {Weiss}}]{Frankfurt:2005mc}%
  \BibitemOpen
  \bibfield  {author} {\bibinfo {author} {\bibfnamefont {L.}~\bibnamefont
  {Frankfurt}}, \bibinfo {author} {\bibfnamefont {M.}~\bibnamefont
  {Strikman}},\ and\ \bibinfo {author} {\bibfnamefont {C.}~\bibnamefont
  {Weiss}},\ }\bibfield  {title} {\bibinfo {title} {{Small-x physics: From HERA
  to LHC and beyond}},\ }\href
  {https://doi.org/10.1146/annurev.nucl.53.041002.110615} {\bibfield  {journal}
  {\bibinfo  {journal} {Ann. Rev. Nucl. Part. Sci.}\ }\textbf {\bibinfo
  {volume} {55}},\ \bibinfo {pages} {403} (\bibinfo {year} {2005})},\ \Eprint
  {https://arxiv.org/abs/hep-ph/0507286} {arXiv:hep-ph/0507286} \BibitemShut
  {NoStop}%
\bibitem [{\citenamefont {Kovner}\ and\ \citenamefont
  {Wiedemann}(2001)}]{Kovner:2001vi}%
  \BibitemOpen
  \bibfield  {author} {\bibinfo {author} {\bibfnamefont {A.}~\bibnamefont
  {Kovner}}\ and\ \bibinfo {author} {\bibfnamefont {U.~A.}\ \bibnamefont
  {Wiedemann}},\ }\bibfield  {title} {\bibinfo {title} {{Eikonal evolution and
  gluon radiation}},\ }\href {https://doi.org/10.1103/PhysRevD.64.114002}
  {\bibfield  {journal} {\bibinfo  {journal} {Phys. Rev. D}\ }\textbf {\bibinfo
  {volume} {64}},\ \bibinfo {pages} {114002} (\bibinfo {year} {2001})},\
  \Eprint {https://arxiv.org/abs/hep-ph/0106240} {arXiv:hep-ph/0106240}
  \BibitemShut {NoStop}%
\end{thebibliography}%

\end{document}